\documentclass[11pt]{article}

\usepackage[utf8]{inputenc}
\usepackage[left=25mm,right=25mm,top=25mm,bottom=25mm]{geometry}

\newcommand{\Softplus}{\mathscr{S\! P}}
\newcommand{\Relu}{\mathscr{R\! L}}
\newcommand{\tr}{\operatorname{tr}}
\newcommand{\norm}[1]{\left\lVert#1\right\rVert}

\newcommand{\red}[1]{\textcolor{black}{#1}}

\usepackage{mathtools}		
\mathtoolsset{centercolon}	

\newcommand{\Sym}{\mathscr{S\! y\! m}} 
\newcommand{\Orth}{\mathscr{O}}
\newcommand{\SO}{\mathscr{S\!O}}
\newcommand{\GL}{\mathscr{G\!\!L}}

\usepackage{upgreek}

\newcommand{\ve}[1]{\boldsymbol{#1}}
\newcommand{\te}[1]{\boldsymbol{#1}}

\newcommand{\diffp}[2]{\frac{\partial #1}{\partial #2}}

\DeclareMathOperator{\trace}{tr}

\DeclareMathOperator{\Cof}{cof}
\newcommand{\cof}{\Cof}

\newcommand{\nablaX}{\nabla_{\!\!{\ve X}}}

\newcommand{\tte}[1]{{\mathbb{#1}}} 

\newcommand{\numbers}[1]{\mathbb{#1}}
\newcommand{\N}{\numbers{N}}    
\newcommand{\R}{\numbers{R}}    
\newcommand{\Lspace}[1]{\mathcal{L}_{#1}}

\usepackage{multirow}
\usepackage{booktabs}
\usepackage{colortbl}

\usepackage{siunitx}  
\usepackage{amssymb}
\usepackage{amsmath}
\usepackage{amsthm}
\usepackage{nicefrac}
\usepackage{bbm}
\theoremstyle{definition}   
\usepackage{chngcntr}
\usepackage{apptools}

\usepackage{mathptmx}      
\usepackage{newtxtext,newtxmath}      

\usepackage[usenames,dvipsnames,svgnames]{xcolor}
\usepackage{subcaption}
\usepackage[font=small]{caption}

\usepackage[style=numeric,
			maxbibnames=9,
			maxcitenames=3,
			natbib=true,
			backend=bibtex,
			isbn=false,
			url=false,
			sortcites
			]{biblatex}
			
\bibliography{literature}

\usepackage[german, main=english]{babel}
\usepackage[autostyle]{csquotes}
\usepackage[T1]{fontenc}
\usepackage{hyperref}
\usepackage[affil-it]{authblk}
\newtheorem{definition}{Definition}[section]
\newtheorem{remark}[definition]{Remark}

\title{Neural networks meet hyperelasticity: A guide to enforcing physics

\vspace{1ex}}

\author[1]{Lennart~Linden}
\author[2]{Dominik~K.~Klein}
\author[1]{Karl~A.~Kalina}
\author[1]{J\"org~Brummund}
\author[2]{Oliver~Weeger}
\author[1,3*]{\;and~Markus~K\"astner}

\affil[1]{\footnotesize Institute of Solid Mechanics, TU Dresden, 01062 Dresden, Germany}
\affil[2]{\footnotesize Cyber-Physical Simulation Group \& Graduate School of Computational Engineering, 
Department of \protect \\  Mechanical Engineering \& Centre for Computational Engineering, TU Darmstadt, 64293 Darmstadt, Germany}
\affil[3]{\footnotesize Dresden Center for Computational Materials Science (DCMS), TU Dresden, 01062 Dresden, Germany}
\affil[*]{\footnotesize Corresponding author, email: markus.kaestner@tu-dresden.de}
\date{May 3, 2023}

\begin{document}
 \maketitle

\par\noindent\rule{\textwidth}{0.4pt}
\begin{abstract}
\noindent
In the present work, a hyperelastic constitutive model based on neural networks is proposed which fulfills all common constitutive conditions by construction, and in particular, is applicable to compressible material behavior.
Using different sets of invariants as inputs, a hyperelastic potential is formulated as a convex neural network, thus fulfilling symmetry of the stress tensor, objectivity, material symmetry, polyconvexity, and thermodynamic consistency.
In addition, a physically sensible stress behavior of the model is ensured by using analytical growth terms, as well as normalization terms which ensure the undeformed state to be stress free and with zero energy.
In particular, polyconvex, invariant-based stress normalization terms are formulated for both isotropic and transversely isotropic material behavior.
By fulfilling all of these conditions in an exact way, the proposed physics-augmented model combines a sound mechanical basis with the extraordinary flexibility that neural networks offer. 
Thus, it harmonizes the theory of hyperelasticity developed in the last decades with the up-to-date techniques of machine learning.
Furthermore, the non-negativity of the hyperelastic \red{neural network-based} potentials is numerically examined by sampling the space of admissible deformations states, which, to the best of the authors' knowledge, is the only possibility for the considered nonlinear compressible models.
For the isotropic neural network model, the sampling space required for that is reduced by analytical considerations.
In addition, a proof for the non-negativity of the compressible Neo-Hooke potential is presented.
The applicability of the model is demonstrated by calibrating it on data generated with analytical potentials, which is followed by an application of the model to finite element simulations.
In addition, an adaption of the model to noisy data is shown and its extrapolation capability is compared to models with reduced physical background.
Within all numerical examples, excellent and physically meaningful predictions have been achieved with the proposed physics-augmented neural network.

\medskip

\end{abstract}
\vspace*{2ex}
\par\noindent\rule{\textwidth}{0.4pt}\vspace*{2pt}
{\small Accepted version of manuscript published in the \emph{Journal of the Mechanics and Physics of Solids}. \\ 
Date accepted: June 14, 2023. DOI: \href{https://doi.org/10.1016/j.jmps.2023.105363}{10.1016/j.jmps.2023.105363}. License: \href{https://creativecommons.org/licenses/by-nc-nd/4.0/legalcode}{CC BY-NC-ND 4.0}}\vspace*{-1.4mm}
\par\noindent\rule{\textwidth}{0.4pt}

\section{Introduction}

The mechanical principles underlying hyperelasticity were extensively discussed in the last decades, but for a long time, fulfilling them all at once could be seen as \enquote{the main open problem of the theory of material behavior} (\textcite{TruesdellNoll}). For instance, while both the polyconvexity \cite{Ball1976,Ball1977} and objectivity condition have a sound mechanical motivation, objective strain measures easily violate the polyconvexity condition \cite{klein2021}. This does not mean that different constitutive conditions contradict each other -- it rather shows the big challenge of fulfilling them all at the same time. 
With an increasing amount of restrictions a model should fulfill, this effort increases considerably, and it took sophisticated approaches to construct analytical models which fulfill all relevant conditions at the same time \cite{Schroeder2003,Ebbing2010}. In addition, the calibration of such models is also not a trivial task and requires a lot of knowledge \cite{Ricker2023}.

To overcome the time consuming task of formulating classical constitutive models and to improve the restricted functional relationships that most of these analytical models bring along, concepts like the data-driven mechanics approach \cite{kirchdoerfer2016} or modern machine learning methods such as Gaussian process regression \cite{Frankel2020,Fuhg2022} or \emph{neural networks (NNs)} \cite{aggarwal2018, kollmannsberger2021} represent promising alternatives. For the first time, the idea of applying NNs in constitutive modeling was proposed in the early 1990s by Ghaboussi~et~al.~\cite{Ghaboussi1991}. However, in this early phase, mostly pure black-box approaches were used, i.e., networks that do not take into account any physical principles and therefore can only reproduce the training dataset, here consisting of stress-strain couples, well but extrapolate poorly. 
To remedy this weakness, a fairly new trend in NN-based constitutive modeling, and in scientific machine learning in general \cite{peng2020}, is to include essential underlying physics in a strong form, e.g., by using adapted network architectures, or in a weak form, e.g., by modifying the loss term for the training \cite{liu2021e}. These types of approaches, coined  as \emph{physics-informed} \cite{karniadakis2021}, \emph{mechanics-informed} \cite{asad2022}, \emph{physics-augmented} \cite{klein2022a}, \emph{physics-constrained} \cite{kalina2022b}, or \emph{thermodynamics-based} \cite{masi2021}, enable an improvement of the extrapolation capability and the usage of sparse training data \cite{kumar2022,karniadakis2021}, which is particularly important when constitutive models are to be fitted to experimental data.

In the following, a brief overview on the mentioned NN-based approaches applied to \emph{finite strain hyperelasticity} modeling is given. 
Regarding isotropic materials, transferred from analytical models, the works \cite{Shen2004,Liang2008} propose to approximate the elastic potential by a feed-forward neural network (FFNN) with three deformation-type invariants as input and thus fulfill several constitutive conditions, e.g., thermodynamic consistency, objectivity, or material symmetry. However, similar to the approaches \cite{Le2015,Ling2016,Sagiyama2019a} applied to anisotropic problems, the elastic potential is needed directly for training within \cite{Shen2004,Liang2008}.
In the meantime, NNs using \emph{invariants} as inputs and the hyperelastic potential as output, thus also being a priori thermodynamically consistent, have become a fairly established approach  \cite{Fuhg2022b,kalina2022b,klein2021,klein2022a,tac2022,Linden2021,Linka2020,linka2023}. Thereby, a more sophisticated training is applied, which allows the direct calibration of the network by tuples of stress and strain, i.e., the derivative of the energy with respect to the deformation is used in the loss term. This technique is also named Sobolev training \cite{Vlassis2020,Vlassis2022a}. Alternatively, in order to ensure thermodynamic consistency a posteriori, a previously trained network predicting stress coefficients can be used to construct a pseudo-potential \cite{kalina2022a}. An NN-based approach which is coupled to a specific model, the so-called micro-sphere approach, is presented in \cite{Zopf2017}. 
Other approaches formulate NN-potentials in terms of the eigenvalues of the deformation gradient \cite{stpierre2023}, thus generalizing Ogden-type models \cite{ogden2004}, or formulate NN-potentials in terms of the components of strain tensors \cite{Fernandez2020,asad2022,klein2021,Vlassis2022a}.

Besides the mentioned requirements, namely thermodynamic consistency, objectivity, and material symmetry, there exist further physical conditions, e.g., ellipticity, which ensures material stability \cite{Zee1983}. However, since ellipticity is difficult to verify and ensure, the concept of polyconvexity of the strain energy potential \cite{Ball1976,Ball1977}, which implies ellipticity and is mathematically linked to the existence and stability of solutions of the elasticity problem, is preferable for the formulation of constitutive models \cite{neff2015}. There are several approaches for building polyconvex NNs \cite{Chen2022,klein2021,linka2023,tac2022,tac2023}, with the most notable technique for incorporating this condition being the use of \emph{input convex neural networks (ICNNs)} originally introduced by Amos~et~al.~\cite{Amos2016}.  
Other approaches consider convexity of the strain energy potential in the right Cauchy-Green deformation tensor $\te C$. 
While, in addition to polyconvexity, this may be a desirable feature of the strain energy potential \cite{lehmich2014}, convexity in $\te C$ alone is not sufficient to ensure material stability \cite{gao2017}.
In the same manner as polyconvex NNs, strain energy potentials which are convex in $\te C$ can be represented with ICNNs \cite{asad2022}. In \cite{Vlassis2020}, convexity of the proposed potential in $\te C$ is examined for one specific dataset. While for the examined case, the proposed potential is indeed convex in $\te C$, this condition is not fulfilled by construction, and, consequently, it may be violated for other applications of the model.
For the fulfillment of the growth condition, a special network architecture may be applied \cite{kalina2022b}, whereas using analytical growth terms is more widely spread \cite{klein2021,klein2022a}. 
Finally, while several works introduce correction terms which ensure normalization conditions, they are either not polyconvex~\cite{thakolkaran2022,Fernandez2020,asad2022} or restricted to the case of nearly incompressible material behavior~\cite{linka2023}. E.g., the model proposed in~\cite{linka2023} includes terms of the form $(I_1-3)^2$, where $I_1$ denotes the first invariant of the right Cauchy-Green deformation tensor. However, in order to preserve polyconvexity of $I_1$, the functions acting on it must in general be convex and \emph{non-decreasing}, cf.~\cite[Remark~A.10]{klein2021}. For the quadratic function used in \cite{linka2023}, this holds only if $I_1$ is bounded from below by 3, which is only the case for $\det \te F=1$, cf.~\cite[Corollary~A.11]{Schroeder2003}, where $\te F$ denotes the deformation gradient. Thus, for $\det \te F \neq 1$, the polyconvexity of the potential in \cite{linka2023} is not ensured by construction.  
Overall, to the best of our knowledge, the models found in literature so far only fulfill subsets of constitutive conditions belonging to compressible hyperelasticity in an exact way, while the remaining ones are only fulfilled in an approximate fashion, i.e., they are taken into account by penalty terms in the loss \cite{Tac2022a,Weber2021}.
Becoming more specific, while \cite{klein2021} fulfills the polyconvexity condition in an exact way, the normalization condition is only approximated by learning it through the calibration data. On the other side, while \cite{thakolkaran2022} uses a stress correction for the exact fulfillment of the normalization condition, this stress correction term includes non-diagonal components of the right Cauchy-Green deformation tensor, and thus is in general not polyconvex and violates material symmetry.

Concluding on NN-based constitutive models, fulfilling all common constitutive conditions of compressible hyperelasticity in an exact way at the same time has so far remained an open challenge.\footnote{This is only true for the case of compressible elastic materials. A corresponding proposal is made by Linka~and~Kuhl~\cite{linka2023} for the incompressible case. However, apart from the restriction to incompressible materials, it is less general in some other points compared to the approach presented here.}
In particular, to the best of the authors' knowledge, no polyconvex stress normalization, which also not violates the balance of angular momentum, objectivity, and material symmetry conditions, has been proposed yet for compressible material behavior.
In the present work, such an approach which consequently accounts for \emph{thermodynamic consistency, symmetry of the stress tensor, objectivity, material symmetry, polyconvexity, growth condition} as well as \emph{normalization of energy and stress} by construction of the network architecture is systematically derived. 
Regarding the introduced new family of NN-based hyperelastic models, which fulfill all of the aforementioned conditions in an \emph{exact} way, we advocate for naming them as \emph{physics-augmented neural networks (PANNs)}. The proposed framework will be very valuable in fields, where highly flexible and at the same time physically sensible constitutive models are required, such as the simulation of microstructured materials \cite{gaertner2021,kalina2022b}.
For this purpose, the PANN approach is build up by extending the aforementioned model \cite{klein2021}. By formulating polyconvex, invariant-based stress normalization terms, the missing link between \emph{polyconvexity} and \emph{stress normalization} in the compressible case is made. As pointed out in Remark~\ref{rem:norm}, these conditions are particularly challenging to combine. 
This is followed by a detailed analytical and numerical study analyzing the overall model. This includes extensive investigations with different model approaches, from a naive NN model which does not include any mechanical conditions, up to the PANN model including all mechanical conditions. Both interpolation and extrapolation of the different models are examined thoroughly, and statistical evaluations are carried out. Thereby, it is clearly explained step by step how all the considered physical conditions are incorporated into the PANN approach, which is particularly applicable to compressible anisotropic material behavior. Our framework is applied to the isotropic and transversely isotropic case, where several descriptive examples including multiaxial stress-strain states and noisy data are considered.
The data basis for training is thereby generated by using analytical potentials. After the calibration of the models, they are applied within finite element (FE) computations to demonstrate their usability and general accuracy. Finally, a proof for the non-negativity of a compressible Neo-Hooke potential is presented, which, to the best of the authors' knowledge, has not been done yet.

The outline of the manuscript is as follows: In Sec.~\ref{sec:basics}, the fundamentals of hyperelasticity are discussed, which are then applied to the proposed NN model in Sec.~\ref{sec:model}. In Sec.~\ref{sec:num}, numerical examples are presented. Finally, in Sec.~\ref{sec:conc} we conclude with a short discussion on the importance of augmenting neural networks with physics. Besides some auxiliary appendices, in App.~\ref{app:non_neg}, a proof for the non-negativity of a compressible Neo-Hooke potential is presented.

\paragraph{Notation}
Throughout this work the space of tensors 
	\begin{align}
		\Lspace{n}:=\underbrace{\R^3 \otimes \cdots \otimes \R^3}_{n\text{-times}} \ \forall n\in\N\label{eq:tensorSpace}
	\end{align}
is used, except for a tensor of rank zero. In Eq.~\eqref{eq:tensorSpace}, $\R^3$, $\N$ and $\otimes$ denote the Euclidean vector space, the set of natural numbers without zero and the dyadic product, respectively. Tensors of rank one and two are given by boldface italic symbols in the following, i.e., $\ve a \in \Lspace{1}$ or $\te B,\te C \in \Lspace{2}$. 
Transpose and inverse of a second order tensor $\te B$ are marked by $\te B^T$ and $\te B^{-1}$, respectively. Furthermore, trace, determinant and cofactor are denoted by $\trace \te B$, $\det \te B$ and $\cof \te B := \det(\te B) \te B^{-T}$.
The set of invertible second order tensors with positive determinant is denoted by $\GL^+(3):=\left\{\te A \in \Lspace{2}\,|\,\det \te A > 0\right\}$, while the orthogonal group and special orthogonal group in $\R^3$ are denoted by $\Orth(3):=\left\{\te A \in \Lspace{2}\,|\,\te A^T \cdot \te A = \te 1\right\}$ and $\SO(3):=\left\{\te A \in \Lspace{2}\,|\,\te A^T \cdot \te A = \te 1,\,\det \te A = 1\right\}$, respectively. Here, $\te 1\in \Lspace{2}$ denotes the second order identity tensor. 
The space of symmetric second order tensors is denoted as \mbox{$\Sym:=\left\{\te A \in \Lspace{2} \, |\, \te A = \te A^T\right\}$}.
Furthermore, the single and double contraction of two tensors are given by $\te B \cdot \te C = B_{kq} C_{ql} \ve e_k \otimes \ve e_l$ and $\te B:\te C=B_{kl}C_{lk}$, respectively. Thereby, $\ve e_k\in \Lspace{1}$ denotes a Cartesian basis vector and the Einstein summation convention is used.
$\nablaX$ is the nabla operator with respect to reference configuration $\mathcal{B}_0$.
For reasons of readability, the arguments of functions are usually omitted within this work. However, to show the dependencies, energy functions are given with their arguments, except when derivatives are written. Also, we let the symbol of a function be identical to the symbol of the function value itself.

\section{Fundamentals of hyperelasticity}\label{sec:basics}

\subsection{Kinematics, stress measures and balance equations}
\label{sec:kinematic_stress}
Let us consider the motion of a solid body, which in its reference configuration  at the time $t_0 \in \R$ is given by \mbox{$\mathcal{B}_0 \subset \R^3$}. After deformation, the current configuration of the body is given by \mbox{$\mathcal{B} \subset \R^3$} at time \mbox{$t\in \mathcal T:=\{\tau\in\R \,|\,\tau \ge t_0\}$}. The motion of the body is defined by a bijective mapping $\ve \varphi: \mathcal{B}_0 \times \mathcal T \to \mathcal{B}$, linking material particles $\ve X\in\mathcal{B}_0$ to $\ve x=\ve \varphi\left(\ve X, t\right) \in \mathcal{B}$. Associated with $\ve \varphi$, the deformation gradient $\te F \in \GL^+(3)$ and its determinant are defined as $\te F := (\nablaX \ve \varphi)^T$ and \mbox{$J:=\det \te F \in \R_{>0}$}. A deformation measure, which is free of rigid body motions, is given by the positive definite right Cauchy-Green deformation tensor \mbox{$\te C:=\te F^T \cdot \te F \in \Sym$}.

Following the concepts of nonlinear continuum mechanics, we introduce the symmetric Cauchy stress \mbox{$\te \sigma \in \Sym$} as well as the first and second Piola-Kirchhoff stress tensors $\te P \in \Lspace{2}$ and $\te T \in \Sym$. The latter two follow from the pull back operations \mbox{$\te P := J \te \sigma \cdot \te F^{-T}$} and \mbox{$\te T := \te F^{-1} \cdot \te P$}, respectively.

To complete the set of material independent equations, relevant balance laws are given in short. Accounting for \emph{conservation of mass} given by $\varrho_0 = J \varrho$ with $\varrho_0$ and $\varrho$ denoting the mass densities of reference and current configuration, the \emph{balance of linear momentum} with respect to $\mathcal B_0$ is given by 
\begin{align}
	\nablaX \cdot \te P^T + \varrho_0\ve f = \ve 0  \;	. \label{eq:BalanceMom}
\end{align}
In the equation above, inertia terms are neglected and $\ve f\in\Lspace{1}$ denotes a mass specific force density. 
Furthermore, considering the balances of mass and linear momentum \eqref{eq:BalanceMom}, the \emph{balance of angular momentum} follows to
\begin{align}
	\te P \cdot \te F^T = \te F \cdot \te P^T \; . 
	\label{eq:balanceAngularMomentum}
\end{align}
Consequently, by using the stress transformations given above, one finds that the conservation of angular momentum requires the symmetry of the stress tensors $\te \sigma$ and $\te T$.
The introduced mechanical balance equations are complemented by the boundary conditions $\ve u = \hat{\ve u}$ on $\partial\mathcal B_0^{\ve u}$ and $\ve p = \hat{\ve p}$ on $\partial \mathcal B_0^{\ve p}$, where $\partial\mathcal B_0^{\ve u}$ and $\partial \mathcal B_0^{\ve p}$ denote essential and natural boundaries for which $\partial\mathcal B_0^{\ve u} \cup \partial\mathcal B_0^{\ve p}=\partial \mathcal{B}_0$ and $\partial\mathcal B_0^{\ve u} \cap \partial\mathcal B_0^{\ve p}=\varnothing$ hold.
Finally, by taking into account the previously introduced balance equations as well as the \emph{balances of energy} and \emph{entropy}, it follows the \emph{Clausius-Duhem inequality} as a consequence of the \emph{second law of thermodynamics}. If thermal effects are neglected, this relation is given by
\begin{align}
	-\dot W +  \te P : \dot{\te F}^T \ge 0 \; , \label{eq:CDU}
\end{align}
where $W$ denotes the Helmholtz free energy density with respect to $\mathrm d V_0 \subset \mathcal B_0$ and $\dot{(\cdot)}$ is the material time derivative.

\subsection{General requirements for hyperelasticity}
\label{subsec:requirements}
\begin{figure}[t!]
	    \centering
	    \includegraphics[width=\textwidth]{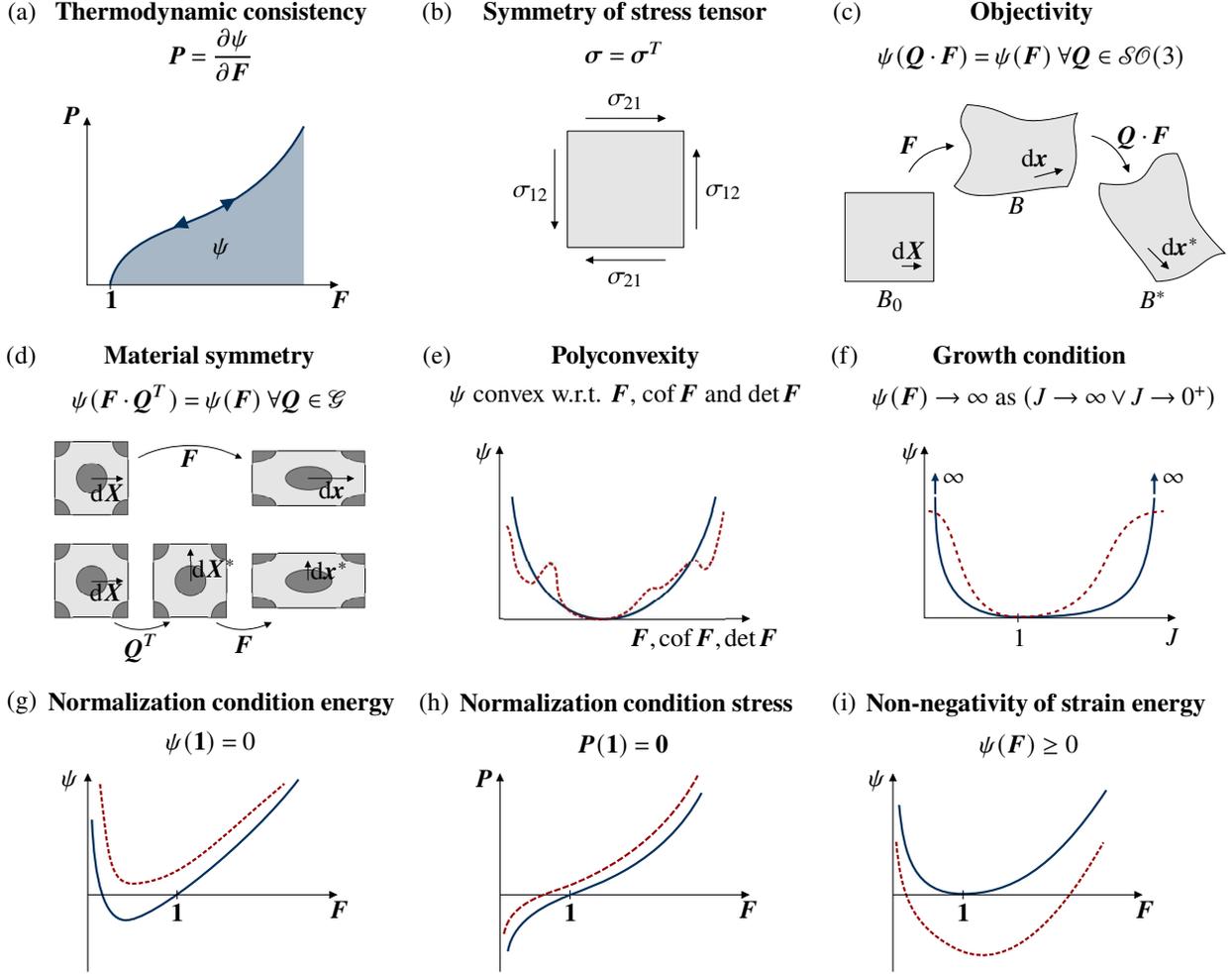}
	    \caption{Schematic depiction of the common conditions on elastic potential and stresses: (a) thermodynamic consistency, (b) symmetry of the Cauchy stress, (c) objectivity, (d) material symmetry, (e) polyconvexity, (f) volumetric growth condition, (g) and (h) normalization conditions for energy and stress, as well as (i) non-negativity of energy. Within (e)--(i), solid blue lines denote models accounting for the respective condition, while the dashed red lines mark models which violate it. In (c) and (d), $\mathrm d \ve X$, $\mathrm d \ve X^*$, $\mathrm d \ve x$, and $\mathrm d \ve x^*$ denote material line elements for different deformation states, where $\mathrm d \ve X^*$ and $\mathrm d \ve x^*$ are transformed by a rotation of $\mathrm d \ve X$ and $\mathrm d \ve x$, respectively.}
	    \label{fig:principles}
\end{figure}
The aim of constitutive modeling in elasticity is to find a connection between the strain at a material point and the stress it evokes.
In hyperelasticity, the mapping between strain and stress is not defined directly, but a potential
\begin{align}
    \psi\colon\GL^+(3)\rightarrow\R,\quad \te F \mapsto \psi(\te F)
    \label{eq:hyperelasticity_energy}
\end{align}
is introduced, which corresponds to the strain energy density stored in the body and is equal to the Helmholtz free energy density $W$.
Thus, to satisfy the inequality~\eqref{eq:CDU} for arbitrary $\dot{\te F}$, the corresponding first Piola-Kirchhoff stress follows to 
\begin{align}
    \te P = \frac{\partial \psi}{\partial \te F}\,.
    \label{eq:hyperelasticity_stress}
\end{align}
By this definition, the stress tensor is a gradient field which implies energy conservation and path-independency, thus being \textbf{thermodynamically consistent} by construction \cite{kalina2022b,Linden2021,linka2023}.\footnote{Note that the constitutive equations must always be physically consistent, i.e., without contradiction to the  introduced balance equations. Special importance is attached to the compatibility with the second law of thermodynamics, given here in the form of the Clausius-Duhem inequality~\eqref{eq:CDU}. Therefore, the thermodynamic consistency is especially highlighted here.} 

%
The hyperelastic potential is subjected to further mathematical and physical considerations, which are shortly discussed in the following and also illustrated in Fig.~\ref{fig:principles}.
For a more detailed introduction to hyperelasticity, the reader is referred to \cite{Haupt2002,Holzapfel2000,Ricker2023}.

\medskip

First of all, $\psi(\te F)$ has to be constructed such that the compatibility with the balance of angular momentum is ensured.
By using Eqs.~\eqref{eq:balanceAngularMomentum} and \eqref{eq:hyperelasticity_stress} this requirement is expressed as 
\begin{align}
    \frac{\partial \psi}{\partial \te F} \cdot \te F^T = \te F \cdot \frac{\partial \psi}{\partial \te F^T} \; .
\end{align}
Using the stress transformations introduced in Sec.~\ref{sec:kinematic_stress}, this results in the requirement for the \textbf{symmetry of the stress tensors} $\te \sigma$ and $\te T$, e.g., $\te T=\te F^{-1} \cdot \frac{\partial\psi}{\partial\te F} \overset{!}{=} \frac{\partial\psi}{\partial\te F^T} \cdot  \te F^{-T} = \te T^T$.

Secondly, a constitutive model should be independent on the choice of observer, which is referred to as \textbf{objectivity}. In hyperelasticity, this is formalized as
\begin{equation}
    \psi(\te Q \cdot \te F) = \psi(\te F)\quad\forall \,\te F \in \GL^+(3),\,\te Q \in \SO(3)\,.
\end{equation}

The constitutive equations should also reflect the material's underlying (an-)isotropy which is expressed as \textbf{material symmetry} and mathematically written as
\begin{equation}\label{eq:mat_symm}
  \psi(\te F \cdot \te Q^T) = \psi( \te F)\quad\forall \,\te F \in \GL^+(3),\,\te Q \in \mathscr{G}\subseteq \Orth(3)\,,
\end{equation}
where $\mathscr{G}$ denotes the symmetry group of the material under consideration. 

Furthermore, we consider hyperelastic potentials which are \textbf{polyconvex} \cite{Ball1976,Ball1977,Ebbing2010,Schroeder2003}, allowing for a representation
\begin{align}
\label{eq:polyconvexity}
    \psi(\te F)=\mathcal P\!\left(\te F,\,\cof \te F,\,\det \te F\right)\,,
\end{align}
where $\mathcal P\!\left(\te F,\,\cof \te F,\,\det \te F\right)$ is convex in its arguments, with $\cof \te F=\det (\te F) \,\te F^{-T}$ denoting the cofactor of the deformation gradient. 
Polyconvexity stems from a quite theoretical context -- however, it implies ellipticity \cite{Zee1983}, which is of practical importance and far more challenging to include in the model formulation than polyconvexity.
The ellipticity (or rank-one convexity) condition \cite{Zee1983, neff2015}
\begin{align}
\left(\ve a \otimes \ve b\right)\colon\frac{\partial^2\psi}{\partial \te F \partial \te F}\colon \left(\ve a \otimes \ve b\right) \geq 0 \qquad \forall \ve a, \ve b \in \Lspace{1}\,,
\end{align}
ensures material stability of the model, which leads to a favorable behavior in numerical applications. Note that, in order to ensure polyconvexity, all steps that are made in the construction of the hyperelastic potential must preserve polyconvexity, in particular, polyconvex invariants have to be used \cite{merodio2006}, cf. Sec.~\ref{sec:spec_aniso}.

In addition, a variety of coercivity conditions can be considered, the most common one being the \textbf{volumetric growth condition}
\begin{align}
    \psi\left(\te F\right) \rightarrow \infty \quad \text{as}\quad \big(J \rightarrow 0^+ \quad\lor\quad J\rightarrow\infty\big)\;,
\end{align}
in order to take into account the observation that a material body can not be compressed to a volume of zero or expanded to an infinite volume~\cite{Holzapfel2000}. 
In particular, the case of large volumetric compression is important, because it may be in a relevant range of practical engineering applications. 

Finally, the functional relationship given in Eq.~\eqref{eq:hyperelasticity_energy} is subjected to further considerations on a physically sensible behavior. In the undeformed configuration, i.e., $\te F = \te 1$, the \textbf{normalization} conditions
\begin{align} \label{eq:norm}
    \psi(\te F = \te 1)\overset{!}{=}0 \quad \text{and} \quad \te P(\te F = \te 1) \overset{!}{=} \te 0
\end{align}
for both energy and stress should hold. 
Besides the normalization, the free energy should increase in any case if deformation appears. Thus, in addition to Eq.~\eqref{eq:norm}${}_1$, the \textbf{non-negativity of the strain energy}, i.e., $\psi(\te F)\geq 0$, is required.

\medskip

By formulating the potential in terms of invariants following from the right Cauchy-Green deformation tensor $\te C$ and a set of structural tensors, summarized in a set $\mathcal S^\square$, which reflects the material symmetry of the material body under consideration \cite{Haupt2002,Holzapfel2000}, i.e.,
\begin{align}\label{psi_inv}
    \psi\colon\R^m\rightarrow\R,\quad \boldsymbol{\mathcal I}\mapsto \psi\left(\boldsymbol{\mathcal I}\right)\,,
\end{align}
both the \textbf{objectivity} and the \textbf{material symmetry} condition are fulfilled \cite{Ricker2023}.\footnote{Note that $\psi(\te F)$ and $\psi(\ve{\mathcal I})$ are different functions, as indicated by the arguments. In favor of a reduced set of symbols, this casual notation is used within this work.} Therein, $\ve{\mathcal I} := (I_1, \dots, I_m) \in\R^m$ denotes an $m$-tuple containing a set of complete and irreducible invariants $I_\alpha(\te C, \mathcal S^\square)$ of the symmetry group under consideration. 
Here, complete means that an arbitrary invariant can be expressed as a function of the elements of $\ve{\mathcal I}$, e.g., $\psi(\ve{\mathcal I})$. 
Irreducible means that one invariant $I_\alpha$ of the set $\ve{\mathcal I}$ cannot be expressed by the other elements of $\ve{\mathcal I}$, cf. \cite{Ebbing2010}.
Note that these invariants have to be chosen such that they do not violate the \textbf{polyconvexity} condition.\footnote{It should be noted that there are symmetry groups for which no complete set can be found that does not violate polyconvexity \cite{klein2022a}. In this case it must be decided which property of the model is preferable. Furthermore, it might be possible that there does not exist a set of invariants to describe the anisotropy of materials with arbitrary microstructures, e.g., 3D-printed composites.} The corresponding second Piola-Kirchhoff stress tensor is then given as
\begin{align}
    \te T= 2\frac{\partial\psi}{\partial\te C} = 2 \sum_{\alpha=1}^m\frac{\partial \psi}{\partial I_\alpha} \frac{\partial I_\alpha}{\partial \te C}=\te F^{-1} \cdot \te P\;.
\end{align}
Finally, formulating the hyperelastic potential in terms of invariants of the right Cauchy-Green tensor implies \textbf{symmetry of the Cauchy stress tensor} $\te \sigma$, which ensures the conservation of angular momentum \cite{Holzapfel2000}.

\subsection{Special material symmetry classes and specific models} \label{sec:spec_aniso}
Throughout this work, both isotropic ($\square := \circledcirc$) and transversely isotropic ($\square := \;\parallel$) material behavior are considered as examples of specific material symmetry groups. In the isotropic case, i.e., when the material response is direction-independent, $\mathscr{G}= \Orth(3)$, a complete and irreducible set is given by the three invariants
\begin{align}
    I_1:=\tr\te C ,\qquad I_2 := \tr (\cof \te C),\qquad I_3 := \det \te C\,.
\end{align}
In this case, there are no structural tensors needed, hence $\mathcal S^\circledcirc = \varnothing$ holds.
For transversely isotropic materials, a complete functional basis is given by the three isotropic invariants $I_1,\,I_2,\,I_3$ together with
\begin{align}
    I_4:=\tr(\te C \cdot\te G),\qquad I_5 := \tr (\cof (\te C)\cdot\te G)\,,
\end{align}
where $\te G$ denotes the second order transversely isotropic structural tensor~\cite{Ebbing2010,Schroeder2008}. In case the preferred direction is parallel to the $X_1$-direction, the structural tensor's components are given by
\begin{align}
\label{eq:structural_tensor}
    (G_{KL}) := \begin{pmatrix}
  \beta^2 & 0 & 0\\ 
  0 & \frac{1}{\beta} & 0\\
  0 & 0 & \frac{1}{\beta}
\end{pmatrix}\, ,
\end{align}
where $\beta\in\R_{>0}$ is a model parameter~\cite{Schroeder2008}. Thus, it holds $\mathcal S^\parallel = \{\te G\}$.

\subsubsection{Isotropic model}
\label{sec:model_NH}

At this point, analytical models have to make an explicit choice of the functional relationship for the hyperelastic potential. While some choices have a strong physical motivation \cite{Martin2017}, most models are of a heuristic nature -- and the reduced flexibility that often goes along with this human choice of functional relationship purely stems from the necessity of an explicit form of the model.
To become more specific, we consider the isotropic Neo-Hooke model 
\begin{align}
\label{eq:energy_NH}
    \psi^{\text{nh}}(I_1,I_3)=\frac{1}{2}\left(\mu\left(I_1-\ln I_3 - 3\right)+\frac{\lambda}{2}\left(I_3-\ln I_3 -1\right)\right)\,, \quad \mu = \frac{E}{2(1 + \nu)}\,, \quad \lambda = \frac{E \nu}{(1 + \nu)(1 - 2\nu)}
\end{align}
with the material parameters $(E,\nu)$ corresponding to the Young's modulus and Poisson's ratio .
From this potential, the second Piola-Kirchhoff stress can be derived as 
\begin{align}
\label{eq:stress_NH}
    \te T^{\text{nh}} = \mu\te 1 + \left(\frac{\lambda}{2}-\frac{2\mu+\lambda}{2I_3}\right)\cof \te C\,.
\end{align}
Note that, e.g., the linear dependency on $I_1$ is a quite restrictive choice.
Again, the second law of thermodynamics is fulfilled by constructing the stress as a gradient field, while symmetry of the Cauchy stress tensor, objectivity and material symmetry are fulfilled by using invariants. 
Then, the normalization conditions are fulfilled by cleverly combining both material parameters and invariants, which becomes evident when setting $ \te C = \te 1,\,I_1=3,\,I_3=1$ in Eqs.~\eqref{eq:energy_NH}~and~\eqref{eq:stress_NH}.
By using the term $-\ln I_3$ in combination with the remaining invariants, the volumetric growth condition is also fulfilled.
Finally, all mathematical operations preserve the polyconvexity of the invariants.

Surprisingly, to the best of the authors' knowledge, the non-negativity of the elastic potential, i.e., $\psi^\text{nh}(I_1,I_3)\ge0$ for arbitrary physically admissible combinations of invariants $I_1$ and $I_3$, has not been shown yet. This is particularly challenging due to the fact that $\psi^\text{nh}(I_1,I_3)$ is polyconvex, i.e., convex w.r.t. $\te F$, $\cof \te F$ and $\det \te F$, but not convex w.r.t. $\te F$ alone or convex w.r.t. $\te C$. \footnote{This is in contrast to the incompressible case \cite{linka2023}, where the definition range of the invariants is significantly more limited since $J\equiv 1$.} 
In particular, the Neo-Hooke potential is not convex in $\te F$ as the invariant $I_3$ is not convex in $\te F$ \cite{smith2018}.
In Theorem~\ref{theorem:neo_hooke}, a proof for the non-negativity of the compressible Neo-Hooke potential is presented.

\subsubsection{Transversely isotropic model}
\label{sec:model_TI}

In order to describe transversely isotropic material behavior, we consider the analytical model
\begin{align}
\label{eq:energy_ti}
    \Tilde \psi^{\text{ti}}(I_1,I_2,I_3,I_4,I_5)=\alpha_1 I_1 + \alpha_2 I_2 + \delta_1 I_3 - \delta_2 \ln(\sqrt{I_3}) + \eta^* \big(I_4^{\alpha_4} + I_5^{\alpha_4}\big)\,, \quad \eta^* := \frac{\eta_1}{\alpha_4 (\tr\te G)^{\alpha_4}}\,,
\end{align}
as proposed by Schr\"oder~et~al.~\cite{Schroeder2008}. 
This transversely isotropic potential presented in Eq.~\eqref{eq:energy_ti} does not fulfill the energy normalization condition, cf.~Eq.~\eqref{eq:norm}.
In order to include this condition we introduce the slightly adapted version
\begin{align}
   \psi^{\text{ti}}(I_1,I_2,I_3,I_4,I_5):=\Tilde\psi^{\text{ti}}(I_1,I_2,I_3,I_4,I_5)+ \Tilde\psi^{\text{ti, energy}}
\end{align}
with the normalization term
\begin{align}\label{eq:ti_energy}
    \Tilde\psi^{\text{ti, energy}}:=-\Tilde\psi^{\text{ti}}(I_1,I_2,I_3,I_4,I_5)\Big\rvert_{\te C = \te 1} = -\left(3\alpha_1 + 3\alpha_2 + \delta_1 + 2\frac{\eta_1}{\alpha_4}\right)\,,
\end{align}
where it already becomes evident that fulfilling the energy normalization condition is straightforward for arbitrary complex strain energy functions. As the energy normalization term Eq.~\eqref{eq:ti_energy} is a constant, it neither influences the polyconvexity of the original potential Eq.~\eqref{eq:energy_ti}, nor its corresponding second Piola-Kirchhoff stress
\begin{align}
\label{eq:stress_ti}
    \te T^{\text{ti}} = 2\bigg(\alpha_1 \te 1 + \alpha_2 \big(I_1 \te 1 - \te C\big) + \Big(\delta_1 I_3 - \frac{\delta_2}{2}\Big) \te C^{-1} + \alpha_4 \eta^* I_4^{\alpha_4 - 1} \te G + \alpha_4 \eta^* I_5^{\alpha_4 - 1} \Big(I_5 \te C^{-1} - \cof(\te C) \cdot \te G \cdot \te C^{-1}\Big)\bigg)\,.
\end{align}
The constitutive model thus includes the material parameters $(\beta, \alpha_1, \alpha_2, \delta_1, \delta_2, \alpha_4, \eta_1)$, with restrictions given in \cite{Schroeder2008}.
Herein, all conditions introduced in Sec.~\ref{subsec:requirements} are satisfied by the analytical model. 
However, in contrast to Sec.~\ref{sec:model_NH} and to the best of the authors' knowledge, the non-negativity of the strain energy density cannot be proven analytically and can thus only be verified numerically.

\medskip

To close, while the constitutive conditions on hyperelasticity discussed in this section have a sound physical and mathematical basis, the restricted functional relationships that most analytical models choose have not. This limitation can be circumvented by using PANNs.

\section{Physics-augmented neural network constitutive model}
\label{sec:model}

In the previous section, we discussed the constitutive conditions of hyperelasticity. 
Now, instead of choosing an analytical formulation for the potential, like in Secs.~\ref{sec:model_NH} and \ref{sec:model_TI}, we aim to exploit the excellent approximation properties of FFNNs \cite{aggarwal2018,kollmannsberger2021}.
In the following, the overall PANN model is introduced, which satisfies all further introduced conditions, except for the non-negativity of the energy, in an exact way, see Sec.~\ref{subsec:requirements}.
\subsection{Basic conditions}
\label{subsec:Basic}
By constructing the stress as the gradient of a potential predicted directly by a FFNN, \textbf{thermodynamic consistency} of the model is fulfilled. If the potential is additionally formulated in terms of invariants, a \textbf{symmetric Cauchy stress tensor}\footnote{It should be noted that a FFNN directly mapping from $\te F$ to $\te P$ does not necessarily yield an elastic model which accounts for $\te \sigma = \te \sigma^T$. The same restriction holds for a model which maps from $\te F$ to $\psi(\te F)$. E.g., in \textcite{Naumann2015}, a pseudo elastic model is investigated which violates the integrability condition, so that the stress is no longer a gradient field.} is automatically obtained and the condition of \textbf{objectivity} as well as the condition of \textbf{material symmetry} are satisfied. At this stage, the neural network already fulfills \emph{basic conditions} by construction, i.e., thermodynamic consistency, objectivity, material symmetry and symmetry of the stress tensor.

We assume that the anisotropic hyperelastic material behavior can be described by the irreducible and independent set of invariants $\ve{\mathcal I} := (I_1, \dots, I_m)$ with $I_\beta(\te C, \mathcal S^\square), \;\beta \in \mathbb{N}_{\leq m}$.
However, it may be necessary to incorporate further invariants $I_{\gamma}^*(\ve{\mathcal I}), \gamma \in \N_{\leq A}$ with $A \in \N_{\geq 0}$ into the argument list of the predicted potential, e.g., in order to fulfill additional physical conditions or to increase the approximation quality of the predictions~\cite{klein2021, kalina2022b}.
By using the extended set of invariants  $\ve{\mathcal I}^* := (I_1, \dots, I_m, I_1^*, \dots, I_A^*)$ as inputs of a FFNN with scalar-valued output which is then taken as a hyperelastic potential $\psi(\ve{\mathcal I}^*)$, the flexibility of the resulting model exceeds the one of analytical formulations by far. 
Furthermore, in a lot of practical applications it is sufficient to restrict the network architecture to only one hidden layer containing  $N^\text{NN}$ neurons. Using the activation function $\mathscr{F}: \R \to \R$, which needs to be continuously differentiable twice, the most simple case of an invariant-based potential using only one hidden layer is given by
	\begin{align}
	\label{eq:psi_NN}
    \psi^{\text{NN},\square} (\ve{\mathcal I}^*) :=  
    \sum_{\alpha=1}^{N^\text{NN}} W_{\alpha}\,\mathscr{F}\left(\sum_{\beta=1}^{m} w_{\alpha\beta} I_\beta + \sum_{\gamma=1}^{A} w_{\alpha\gamma}^* I_\gamma^* + b_\alpha\right)\; ,
	\end{align}
where $W_{\alpha}, w_{\alpha\beta},  w_{\alpha\gamma}^*$ and $b_\alpha$ denote weights and bias values, respectively, which together form the set of parameters $\boldsymbol{\mathcal P} \in \R^P$, with $P\in \N$ denoting the total number of parameters, to be optimized in the calibration process to fit a given dataset.
Note that none of the methods introduced in this paper is restricted to FFNNs with one hidden layer, but can directly be applied for multilayered network architectures, cf. Appendix~\ref{sec:multilayer}.
Nevertheless, we use only one layer here to better illustrate the derivations.
Then, the stress prediction is done by applying 
	\begin{align}
		\te T^{\text{NN},\square} = 2 \sum_{\alpha = 1}^{m} \diffp{\psi^{\text{NN},\square}}{I_\alpha} \diffp{I_\alpha}{\te C}  + 2 \sum_{\gamma = 1}^{A} \sum_{\beta=1}^m \diffp{\psi^{\text{NN},\square}}{I_\gamma^*} \diffp{I_\gamma^*}{I_\beta} \diffp{I_\beta}{\te C}
		\; . \label{eq:StressNN} 
	\end{align}
This special choice of \emph{input quantities} -- namely, invariants -- is the first way of including physics into the NN.

\subsection{Polyconvexity}
For ensuring \textbf{polyconvexity} of the potential, it is necessary to use polyconvex invariants of the symmetry group under consideration.
In addition, the \emph{network architecture} must be adapted in a certain way.
By using a convex and non-decreasing activation function $\mathscr{F}$ and non-negative weights, polyconvexity of the overall potential $\psi^{\text{NN},\square}(\ve{\mathcal I}^*)$ is ensured \cite{klein2021}.
Here, the \emph{Softplus} activation function $\Softplus(x):=\log(1+\exp(x))\in\mathcal{C}^{\infty}$ is applied, which is convex and non-decreasing, which overall leads to the conditions
\begin{align}
	\label{eq:weights_poly}
    W_\alpha, w_{\alpha\beta}, w_{\alpha\gamma}^* \in \R_{\geq 0}, b_\alpha \in\R \quad \forall \alpha\in\N_{\leq N^\text{NN}}, \beta \in \N_{\leq m}, \gamma \in \N_{\leq A}\, .
	\end{align}
\begin{remark}[\textbf{Input convex neural networks (ICNNs)}]\label{rem:ICNN}
This special kind of FFNN, namely, one with its output being convex in its input arguments, is referred to as ICNN \cite{Amos2016}. 
In order to understand what makes this kind of NNs \emph{convex}, we take a step back and consider the univariate function
\begin{equation}
f\colon \R\to \R\,,\quad x\mapsto f(x) := (g\circ h)(x)\,,
\end{equation}
where $f$ is composed of two functions $g,h\colon \R \to \R$.
Given that all of the above functions are twice continuously differentiable, convexity of $f(x)$ in $x$ is equivalent to the non-negativity of its second derivative
\begin{equation}
    f''(x) = (g''\circ h)(x)\, h'(x)^2 + (g'\circ h)(x)\, h''(x) \geq 0\,.
\end{equation}
A sufficient, albeit not necessary condition for this is that the innermost function $h$ is convex $(h'' \geq 0)$, while $g$ is convex and non-decreasing $(g'' \geq 0 \text{ and } g' \geq 0)$. This can be generalized to arbitrarily many function compositions, where the innermost function must be convex, while every following function must be convex and non-decreasing. Transferred to FFNNs, which can be seen as compositions of multiple vector-valued functions, this generalizes to the condition that in an ICNN the first hidden layer must be node-wise convex, while every subsequent layer must be node-wise convex and non-decreasing, cf. \cite[Appendix~A]{klein2021}.

\medskip
When applied to hyperelasticity, the architecture of ICNNs has to be further adapted: as the invariants are nonlinear functions of the arguments defined in the polyconvexity condition, i.e., $(\te F, \cof\te F, \det\te F)$, cf.~Eq.~\eqref{eq:polyconvexity}, they are the innermost function acting on the arguments of the polyconvexity condition. Thus, already the first hidden layer has to be convex and \emph{non-decreasing.}
The only exception is the use of $J=\sqrt{I_3}$ as an additional invariant (instead of $I_3$), since $J$ is an argument of the polyconvexity condition. 
For this reason, the activation function acting on $J$ must only be convex and not necessarily non-decreasing. This is pragmatically taken into account by including the additional invariant $I_1^*:=-2J$ in the set of invariants, which is furthermore essential to represent negative stresses at all~\cite{klein2021}.
Note that there are different ways of constructing polyconvex neural networks, e.g., \cite{tac2022} or \cite{Chen2022}. However, the simple structure and excellent flexibility of ICNNs makes them a very natural choice for this task.
For a more extensive introduction to polyconvex neural networks and explicit proofs, see \cite{klein2021}. 
\end{remark}

\subsection{Growth and normalization conditions}
Finally, \textbf{growth and normalization conditions} remain to be included in the neural network, ensuring a physically sensible stress behavior of the model. 
For this, growth and normalization terms are added to the original potential $ \psi^{\text{NN},\square}(\ve{\mathcal I}^*)$ of Eq.~\eqref{eq:psi_NN}. 
Then, with $\square$ denoting the symmetry group under consideration, the overall PANN model, given as
	\begin{align}
	\label{eq:psi_corr}
    \psi^{\text{PANN},\square}(\ve{\mathcal I}^*) := \psi^{\text{NN},\square}(\ve{\mathcal I}^*) + \psi^{\text{stress},\square}(J,I_4,\dots,I_m) + \psi^{\text{energy},\square} + \psi^{\text{growth}}(J)\, ,
	\end{align}
fulfills all constitutive conditions introduced in Sec.~\ref{subsec:requirements}, except for the non-negativity of the energy, in an exact way. Even for simple analytical models such as the Neo-Hooke model, cf.~Eq.~\eqref{eq:energy_NH}, a proof of the non-negativity of the energy is not straightforward, cf.~App.~\ref{app:non_neg}. Thus, it becomes evident that a proof of non-negativity for fairly general hyperelastic potentials is a daunting task. Consequently, the non-negativity of the PANN model is examined numerically, cf.~Sec.~\ref{sec:num}, where the required sampling space for the isotropic model can be reduced by analytical considerations, cf. Theorem~\ref{theorem:isotropic_PANN}.
\begin{figure}[t!]
	    \centering
	    \includegraphics[width=\textwidth]{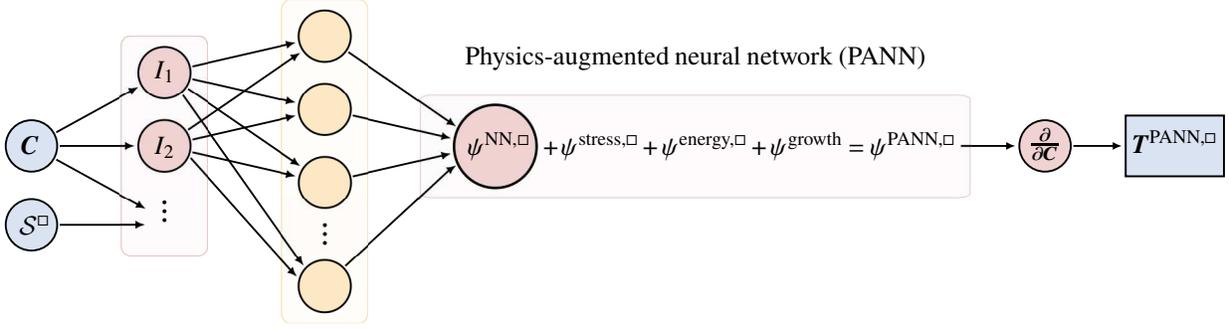}
	    \caption{Illustration of the PANN based constitutive model for the material symmetry group $\square$ under consideration. Note that the hidden-layer (yellow) of the NN may be multilayered.
	    }
	    \label{fig:PANN_scheme_one_hidden_layer}
	\end{figure}
In Fig.~\ref{fig:PANN_scheme_one_hidden_layer}, the overall structure of the PANN model is exemplarily illustrated for one hidden layer.
Consequently, the corresponding stress is given by the expression
	\begin{align}
	\label{eq:stress_corr}
     \te T^{\text{PANN},\square} = 2 \frac{\partial \psi^{\text{PANN},\square}}{\partial \te C} =
    \te T^{\text{NN},\square}   + \te T^{\text{stress},\square} + \te T^{\text{energy},\square}
    + \te T^{\text{growth}}
  \; .
	\end{align}

One way to fulfill the volumetric \textbf{growth condition} are coercive functions. However, since ICNNs are not necessarily coercive, they are not suited to fulfill this condition and therefore an analytical term 
\begin{align}
		\psi^\text{growth}(J) := \Big(J + \frac{1}{J} - 2\Big)^2 \label{eq:GrowthNNmod}
\end{align}
is introduced, which is chosen in such a way that polyconvexity is not violated.
This leads to the corresponding stress contribution
\begin{align}
		\te T^\text{growth} = 2 \Big(J + \frac{1}{J} - 2\Big)\Big(1 -\frac{1}{J^2}\Big) J \te C^{-1}\,. \label{eq:GrowthStressNN}
	\end{align}
Another way to fulfill the volumetric growth condition is further adapting the network architecture, cf. \cite{kalina2022b}, however, using an analytical term is more straightforward.

The correction term for the \textbf{energy normalization} is given by 
\begin{align}
    \psi^{\text{energy},\square}:=-\psi^{\text{NN},\square}(\ve{\mathcal I}^*)\Big\rvert_{\te C = \te 1} \in \R\, .
\end{align}
Since $\psi^{\text{energy},\square}$ is a constant, it holds $\te T^{\text{energy},\square}=\mathbf{0}$. 
Together with the stress normalization term denoted by $\psi^{\text{stress},\square}(J,I_4,\dots,I_m)$, meaning vanishing gradients of the potential for the undeformed state, we ensure 
\begin{align}
    \psi^{\text{PANN},\square}(\ve{\mathcal I}^*)\Big\rvert_{\te C = \te 1} = 0
\end{align}
as well as
\begin{align}
    \te T^{\text{PANN},\square}\Big\rvert_{\te C = \te 1} = 2 \left(\sum_{\alpha = 1}^{m} \diffp{\psi^{\text{PANN},\square}}{I_\alpha} \diffp{I_\alpha}{\te C}  +  \sum_{\gamma = 1}^{A} \sum_{\beta=1}^m \diffp{\psi^{\text{PANN},\square}}{I_\gamma^*} \diffp{I_\gamma^*}{I_\beta} \diffp{I_\beta}{\te C}\right)\Bigg\rvert_{\te C = \te 1} = \te 0\; .
\end{align}
Thus, the potential has a local minimum exactly for the undeformed state with $\te C = \te 1$, where for this case we define the invariants as $I_\alpha^0 := I_\alpha(\te C = \te 1)$ and $I_\beta^{*0} := I_\beta^{*}(\te C = \te 1)$. 
However, due to the fact that $\psi^{\text{PANN},\square}(\ve{\mathcal I}^*)$ is polyconvex, i.e., convex w.r.t. $\te F$, $\cof \te F$ and $\det \te F$, but not convex w.r.t. $\te F$ or $\te C$, the \textbf{non-negativity of the energy}, i.e., $\psi^{\text{PANN},\square}(\ve{\mathcal I}^*)  \ge 0$ does not automatically follow. As already stated in Sects.~\ref{sec:model_NH} and \ref{sec:model_TI}, a numerical test for admissible deformation states is needed to prove the fulfillment of this condition, see also Sec.~\ref{sec:num}.

The polyconvex stress correction $\psi^{\text{stress},\square}(J,I_4,\dots,I_m)$ depends on the symmetry group under consideration and is now further discussed.
\subsubsection{Isotropic normalization term}
In the isotropic case with $\ve{\mathcal I}^* := (I_1, I_2, I_3, I_1^*)$ the normalization term is given by
\begin{align}
\label{eq:normalization_isotropic}
    \psi^{\text{stress,} \circledcirc}(J) := -\mathfrak{n}(J-1)\;,
\end{align}
where the constant
\begin{align}
\label{eq:normalization_A_iso}
\mathfrak{n} := 2 \,\bigg(\,\frac{\partial \psi^{\text{NN},\circledcirc}}{\partial I_1}
 +
 2\frac{\partial \psi^{\text{NN},\circledcirc}}{\partial I_2}
+
 \frac{\partial \psi^{\text{NN},\circledcirc}}{\partial I_3}
+\frac{\partial \psi^{\text{NN},\circledcirc}}{\partial I_1^*} \frac{\partial I_1^*}{\partial I_3}
 \bigg)\Bigg\rvert_{\te C = \te 1} \in \R
\end{align}
is a weighted sum of derivatives of the ICNN potential with respect to the invariants for the undeformed state $\te C = \te 1$. 
The corresponding stress contribution is given by
    \begin{align}
		\te T^{\text{stress,} \circledcirc} = - \mathfrak{n} J \te C^{-1}\; . \label{eq:normalization_StressNN_iso}
	\end{align}
and leads to the fact that the undeformed state is stress-free by construction.

This definition of the correction term comprises two ideas: first of all, this approach preserves polyconvexity of the potential, since the additional term is a linear function in $J$, which is an invariant quantity included in the arguments of the polyconvexity condition, cf. Eq.~\eqref{eq:polyconvexity}. Furthermore, the partial derivatives
    \begin{align}
    \label{eq:tensor_generators_iso}
	\frac{\partial I_\alpha}{\partial \te C}\bigg\vert_{\te C = \te 1}
	= \xi \te 1 \quad \text{and} \quad \frac{\partial I_1^*}{\partial \te C}\bigg\vert_{\te C = \te 1}
	= \eta \te 1
	\end{align}
of the isotropic invariants w.r.t. $\te C$ for the undeformed state are multiples of the identity tensor $\te 1$ for all $\alpha \in \N_{\leq 3}$ with constants $\xi, \eta \in \R$. 
Hence, it can normalize the stress for the undeformed state to zero, which becomes evident when setting $\te C= \te 1$ in Eq.~\eqref{eq:normalization_StressNN_iso}.
Further details are provided in Appendix~\ref{sec:derivatives_inv}, where all necessary tensor derivatives are given in analytical form.

\subsubsection{Transversely isotropic normalization term}

In the case of transverse isotropy with $\ve{\mathcal I}^* := (I_1, I_2, I_3, I_4, I_5, I_1^*)$ the partial derivatives of the additional invariants $I_4,I_5$ w.r.t. $\te C$ for the undeformed state are not multiples of the identity anymore, but include the structural tensor $\te G$, e.g., 
    \begin{align}
    \label{eq:tensor_generators_ti}
	\frac{\partial I_4}{\partial \te C}\bigg\vert_{\te C = \te 1}
	= \te G \quad \text{and} \quad \frac{\partial I_5}{\partial \te C}\bigg\vert_{\te C = \te 1}
	= I_5\te 1 - \te G\;.
	\end{align}
Thus, the correction term 
\begin{align}
    \label{eq:psi_stress_ti}
    \psi^{\text{stress,}\parallel}(J,I_4,I_5) := -\mathfrak{o}(J-1) + \mathfrak{p}(I_4 - I_4^0) + \mathfrak{q}(I_5-I_5^0)
\end{align}
is introduced, with $I_4^0=I_5^0=\tr \te G$. Again, the constant
\begin{equation}
    \begin{aligned}
\label{eq:normalization_A_ti}
    \mathfrak{o}:=2\left(
    \frac{\partial \psi^{\text{NN},\parallel}}{\partial I_1}
+ 2   \frac{\partial \psi^{\text{NN},\parallel}}{\partial I_2}
+    \frac{\partial \psi^{\text{NN},\parallel}}{\partial I_3}
+    \frac{\partial \psi^{\text{NN},\parallel}}{\partial I_1^*} \frac{\partial I_1^*}{\partial I_3}
+    \frac{\partial \psi^{\text{NN},\parallel}}{\partial I_5}\tr\te G
+\mathfrak{q}\,\tr\te G
    \right)\bigg\rvert_{\te C=\te 1} \in \R
    \end{aligned}
\end{equation}
is a weighted sum of the derivatives of the ICNN potential with respect to the invariants for the undeformed state $\te C = \te 1$.
Furthermore, with the ReLU-function denoted as $\Relu$, the non-negative constants
\begin{equation}
\label{eq:normalization_B_D_ti}
    \mathfrak{p}:=\Relu(-x) \in \R_{\geq 0}, \quad \mathfrak{q}:=\Relu(x) \in \R_{\geq 0}
\end{equation}
are defined with the argument
\begin{equation}
\label{eq:normalization_x_ti}
    x:=\left( \frac{\partial \psi^{\text{NN},\parallel}}{\partial I_4}- \frac{\partial \psi^{\text{NN},\parallel}}{\partial I_5}\right)\bigg\rvert_{\te C=\te 1}\,.
\end{equation}
Due to the non-negativity of $\mathfrak{p}, \mathfrak{q}$ as well as the polyconvexity of $I_4, I_5$, the correction term is again polyconvex.
It should be emphasized again that in the correction terms no assumptions on the number of hidden layers in the neural networks are made, and the approach can directly be applied to multilayered network architectures.
Overall, in the transversely isotropic case, the stress contribution of the normalization term is given by
\begin{equation}
    \begin{aligned}
    \label{eq:normalization_StressNN_ti}
    \te T^{\text{stress,}\parallel}
    =
    - \mathfrak{o}J\te C^{-1}
    + 2\mathfrak{p}\te G
    + 2\mathfrak{q}\Big(I_5\te C^{-1}- \cof(\te C) \cdot\te G \cdot \te C^{-1}\Big)\;,
    \end{aligned}
\end{equation}
where all necessary tensor derivatives of the Appendix~\ref{sec:derivatives_inv} are integrated.

Note that in the definitions of all correction terms, even though $\psi^{\text{stress},\square}(J,I_4,I_5)$ and $\psi^{\text{energy}}$ depend on the evaluation of $\psi^{\text{NN},\square}(\ve{\mathcal I}^*)$ and its partial derivatives at $\te C=\te 1$, no assumptions on the number of hidden layers in the ICNN $\psi^{\text{NN},\square}(\ve{\mathcal I}^*)$ are made, and the approaches can directly be applied to multilayered network architectures.
\begin{remark}\label{rem:norm}
Stress normalization terms as proposed in this section again exemplify the challenge of fulfilling several mechanical conditions at once. Without polyconvexity, formulating stress normalization terms is straightforward. In this case, projection approaches such as
\begin{equation}\label{eq:proj_1}
    \Tilde{\psi}^{\text{stress},\square,\text{C}}=-\frac{\partial \psi^{\text{NN},\square}}{\partial \te C}\bigg\vert_{\te C = \te 1}:\left(\te C-\te 1\right)
\end{equation}
can be applied, as proposed by \cite[Eq.~(20)]{Fernandez2020}, see also \cite{asad2022,huang2022} and \cite{thakolkaran2022}, where a slightly adapted version of Eq.~\eqref{eq:proj_1} is applied.
Unfortunately, as not all components of $\te C$ are convex in $\te F$, this approach is not polyconvex. Also, as the correction term is formulated directly in $\te C$, it does not fulfill the material symmetry condition, cf.~Eq.~\eqref{eq:mat_symm}.
A polyconvex counterpart of Eq.~\eqref{eq:proj_1} is given by
\begin{equation}
    \Tilde{\psi}^{\text{stress},\square,\text{F}}=-\frac{\partial \psi^{\text{NN},\square}}{\partial \te F}\bigg\vert_{\te F = \te 1}:\left(\te F-\te 1\right)\,.
\end{equation}
However, as it is formulated directly in $\te F$, this approach does not fulfill the objectivity condition, the balance of angular momentum and the material symmetry condition, cf.~Sec.~\ref{subsec:requirements}. 
In a similar manner, a projection approach can be formulated in terms of the invariants. For the model proposed in this work, this corresponds to
\begin{align}
     \Tilde{\psi}^{\text{stress},\square,\text{I}}= -\sum_{\beta=1}^{m} 	\frac{\partial \psi^{\text{NN},\square}}{\partial I_{\beta}}\bigg\vert_{\te C = \te 1} \, \left(I_\beta-I_\beta^0\right) - \sum_{\gamma=1}^{A}	\frac{\partial \psi^{\text{NN},\square}}{\partial I_{\gamma}^*}\bigg\vert_{\te C = \te 1} \, \left(I_\gamma^*-I_\gamma^{*0}\right)  \; .
\end{align}
This projection approach ensures that the partial derivatives of the adapted potential
\begin{align}
    \Tilde \psi^{\text{NN},\square}:=\psi^{\text{NN},\square}+ \Tilde{\psi}^{\text{stress},\square,\text{I}}
\end{align}
w.r.t. each individual invariant vanish at the identity $\te C = \te 1$, thus ensuring a stress-free reference configuration.
This, in turn, fulfills both objectivity and material symmetry. But again, the invariant-based projection approach does not fulfill the polyconvexity condition, as the subtraction operation required here does not preserve the polyconvexity of the invariants.
\end{remark}

Closely related to the invariant-based projection approach, it is possible to directly formulate the hyperelastic potential so that its gradients w.r.t. the invariants vanish in the reference configuration \cite{tac2022,linka2023}. For the model proposed in this work, this corresponds to
    \begin{align}\label{eq:derivs_equal}
	\frac{\partial \psi^{\text{NN},\square}}{\partial I_{\alpha}}\bigg\vert_{\te C = \te 1}
	= 0 \quad \text{and} \quad \frac{\partial \psi^{\text{NN},\square}}{\partial I_\gamma^*}\bigg\vert_{\te C = \te 1}=0\quad\forall \,\alpha\in\mathbb{N}_{\leq m},\,\gamma\in\mathbb{N}_{\leq A}
	\,.
	\end{align}
In \cite{linka2023}, this is done by applying activation functions such as
\begin{align}\label{eq:act_quad}
    f(I_1)=(I_1-3)^2\,,
\end{align}
whose gradient
\begin{align}
    \frac{\partial f(I_1)}{\partial I_1}=2(I_1-3)
\end{align}
vanishes at the identity $\te C=\te 1,\, I_1=3$. 
However, as pointed out in the introduction, \cite{linka2023} considers incompressible material behavior, which allows for the usage of simple activation functions such as quadratic functions, cf.~Eq.~\eqref{eq:act_quad}. In the compressible case, the invariants do not have a global minimum at the identity $\te C=\te 1$, suitable activation functions become fairly complex. The construction of such functions becomes particularly challenging, as besides a vanishing gradient at some specific point, the function should fulfill other conditions. E.g., it should be convex and monotonically increasing, be at least twice continuously differentiable and should not have a too high slope.
Overall, one candidate for $\alpha\in (0, \infty)$ is the fairly complicated function
\begin{align}
       f(x) = \begin{cases}
        0 & \text{for } x < 0\;,\\
        \frac{\alpha}{6} x^3 - \frac{1}{12} x^4 & \text{for } 0\leq x \leq \alpha\;,\\
        \frac{\alpha^3}{6}x-\frac{\alpha^4}{12} & \text{for } \alpha < x\;.
        \end{cases}
\end{align}
Furthermore, while being suitable for the construction of polyconvex models, this approach is quite restrictive, as it does not even allow for the representation of simple analytical models such as the Neo-Hooke model, cf.~Eq.~\eqref{eq:energy_NH} and reference~\cite{tac2022}. In \cite{linka2023}, this problem is circumvented by assuming perfect incompressibility and applying the hydrostatic pressure as an additional unkown. However, this is not possible for the general, compressible case, which suggests that such methods are limited to the incompressible case.

Yet another approach for transverse isotropy is to formulate the potential such that
\begin{align}\label{eq:norm_equal}
   \frac{\partial \psi^{\text{NN},\parallel}}{\partial I_4}\equiv\frac{\partial \psi^{\text{NN},\parallel}}{\partial I_5}
\end{align}
holds for each deformation state $\te C \in \Sym$.
Actually, Eq.~\eqref{eq:norm_equal} is similar to stress normalization approaches for analytical constitutive models, cf.~\cite[Sec.~3.1]{Schroeder2008}, \cite[Sec.~6.4.1]{Ebbing2010} and Eq.~\eqref{eq:energy_ti}.
Within the framework proposed in the present work, Eq.~\eqref{eq:norm_equal} would fairly faciliate the stress normalization term, cf.~Eqs.~(\ref{eq:psi_stress_ti},\ref{eq:normalization_x_ti}). By that, the transversely isotropic stress normalization term, cf.~Eq.~\eqref{eq:psi_stress_ti}, would take a much easier form similar to the isotropic case, cf.~\eqref{eq:normalization_isotropic}. In the context of NNs, Eq.~\eqref{eq:norm_equal} could be fulfilled by setting all weights and activation funtions equal which act on the $I_4$ and $I_5$ input. However, this would be quite restrictive, as it would globally restrict the functional dependency of the hyperelastic potential in $I_4$ and $I_5$.

In contrast to the just discussed approaches, the invariant-based normalization terms proposed in this work only make minor, less restrictive modifications to the strain energy potential, cf.~Remark~\ref{rem:calib}, while fulfilling all common constitutive conditions by construction.

\subsection{Model calibration}
\label{subsec:model_calibration}

Finally, the model has to be calibrated to data of a specific material including the set of structural tensors, summarized in $\mathcal S^\square$, which corresponds to the material symmetry group $\square$ under consideration. Throughout this work, datasets of the form
\begin{align}
    \mathcal D=\big\{
    \big(\,{}^1\te C,\,{}^1\te T\,\big),\,\big(\,{}^2\te C,\,{}^2 \te T\,\big),\,\dotsc
    \big\}
\end{align}
consisting of strain-stress tuples in terms of right Cauchy-Green deformation and second Piola-Kirchhoff stress tensors are used. In order to examine the generalization of a model, i.e., its prediction of general load cases, it is essential to evaluate it on data not seen in the calibration process. Thus, the overall dataset $\mathcal{D}$ is split into a calibration dataset $\mathcal{D}_\text{c}$ and a test dataset $\mathcal{D}_\text{t}$ with $\mathcal{D}_\text{c}\cap\mathcal{D}_\text{t}=\varnothing$. Then, after the calibration of the model on the dataset $\mathcal{D}_\text{c}$, its predictions can be evaluated on $\mathcal{D}_\text{t}$. Only when the model is able to predict the load cases of the test dataset, and only if the load cases included in the test dataset are sufficiently general, it can be assumed that the model generalizes well and can predict the stress for arbitrary deformations. 

The next step of the model calibration is the choice of its hyperparameters, here, the number of hidden layers and the nodes they contain. For this, a \enquote{sufficiently large} number of layers and nodes should be chosen, so that the model is flexible enough for the material behavior under consideration. 
For general NNs, overfitting \cite{aggarwal2018,kollmannsberger2021}, i.e., a too accurate interpolation of the calibration data which results in a bad prediction of general data, is an issue. However, as we will demonstrate, for the PANN proposed in this work, the inclusion of physics provides the model with a pronounced mathematical structure, which makes it less prone to overfitting.
Then, with a fixed model architecture, the model parameters $\boldsymbol{\mathcal P}$ must be optimized, i.e., its weights and biases.
To calibrate the model parameters $\boldsymbol{\mathcal P}$ on the dataset $\mathcal{D}_\text{c}$, the loss function defined in terms of the mean squared error
\begin{align}
\label{eq:mse_calibration}
    \mathscr{M\!S\!E}^{\square}(\boldsymbol{\mathcal P})=\frac{1}{\vert\mathcal{D}_\text{c}\vert}\sum_{i = 1}^{\vert\mathcal{D}_\text{c}\vert} \Big\Vert{}^i\te T-\te T^{\text{model},\square}({}^i\te C;\,\boldsymbol{\mathcal P})\Big\Vert^2
\end{align}
is minimized, where $\vert\mathcal{D}_\text{c}\vert$ denotes the number of tuples in $\mathcal{D}_\text{c}$ and $\norm{\cdot}$ denotes the Frobenius norm. 
Note that the hyperelastic potential $\psi^\text{PANN}(\ve{\mathcal I}^*)$ is calibrated only through its gradients, i.e., the corresponding stress tensor, which is referred to as Sobolev training \cite{vlassis2021}. 
For the optimization process, the \emph{SLSQP optimizer} (Sequential Least Squares Programming) is applied \cite{Nocedal2006}.
According the proposed model, loss function and optimizer have no influence on the underlying physics, and they could also be chosen otherwise.
An implementation of the described workflow is realized using \emph{Python, TensorFlow} and \emph{SciPy}.

\begin{remark}\label{rem:calib}
In the calibration process, it is important to already account for the growth term $\psi^{\text{growth}}(J)$ included in the model $\psi^{\text{PANN},\square}(\ve{\mathcal I}^*)$, cf. Eq.~\eqref{eq:psi_corr}, as in particular for calibration data including high volumetric compressions, it can have a considerable influence on the model. On the other side, the \emph{stress and energy normalization} terms may not necessarily have to be included in the calibration process: the energy normalization term $\psi^{\text{energy}}$ can be added to the PANN model after calibration, as it is a constant it does not influence the stress prediction at all. Then, even when the model is calibrated without the stress correction term $\psi^{\text{stress},\square}(J,I_4,\ldots,I_m)$, it can learn through suitable data how to approximate the stress-free reference configuration \cite{klein2021}. Given that the approximation of the stress-free reference configuration is good enough, 
the constants $\mathfrak{n}, \mathfrak{o}, \mathfrak{p}$ on which the stress correction depends become very small, cf. Eqs.~(\ref{eq:normalization_A_iso},\ref{eq:normalization_A_ti},\ref{eq:normalization_B_D_ti}). Then, both the correction term $\psi^{\text{stress},\square}(J,I_4,\ldots,I_m)$ and its gradient $\te T^{\text{stress},\square}$ approximately vanish, and adding the stress correction after the model calibration has a negligibly small influence on the overall model behavior. 
While it is indeed possible to fulfill the normalization conditions in good approximation by learning them through suitable data, they are only fulfilled exactly when incorporated in the model formulation, e.g., by normalization terms. 
\end{remark}

\section{Numerical examples}\label{sec:num}

After the introduction of the PANN hyperelastic constitutive model in the former section, the ability of our approach is now demonstrated by several numerical examples and compared to approaches with reduced physical foundation. We start with simple stress states and analyze interpolation and extrapolation behavior, also for perturbed data. Thereafter, we go on to complex multiaxial deformation states. Finally, the trained PANN model is applied within a finite element simulation.

\subsection{Simple stress-strain states}
\label{subsec:uniaxial_tension}
In this subsection, three types of models, namely a network $\te P^{\text{simple}}(\te F)$ directly mapping from $\te F$ to $\te P$, a network $\psi^{\text{NN,}\circledcirc}(\ve{\mathcal I}^*)$ accounting for the basic conditions, and the PANN $\psi^{\text{PANN,}\circledcirc}(\ve{\mathcal I}^*)$, are compared with respect to their interpolation and extrapolation capabilities for simple stress-strain states. Thereby, the first model, given by
\begin{align}
	\label{eq:P_NN}
    P_{kL}^{\text{simple}}(\te F) :=  
    B_{kL} + \sum_{\alpha=1}^{N^\text{NN}} W_{\alpha kL}\,\Softplus\left(\sum_{i=1}^{3}\sum_{J=1}^3 w_{\alpha iJ} F_{iJ} + b_\alpha\right)\; \text{with } B_{kL}, W_{\alpha kL}, w_{\alpha iJ}, b_\alpha \in \R \; ,
\end{align}
does not take into account any of the introduced conditions, the second, which is defined by Eq.~\eqref{eq:psi_NN}, respects for thermodynamic consistency, symmetry of $\te \sigma$, objectivity and material symmetry, and the PANN fulfills the conditions of the previous model and additionally includes polyconvexity, growth condition, as well as energy and stress normalization, see Eq.~\eqref{eq:psi_corr}.

In the following examples, the architectures of the three models are set to one hidden layer with $N^\text{NN} := 4$ neurons. Furthermore, the softplus activation function is used for all models. The calibration of the model accounting for basic conditions and the PANN is performed according to Sec.~\ref{subsec:model_calibration}. In contrast, naturally, the training of the $\te F$-$\te P$ model is performed with data for $\te F$ and $\te P$. Furthermore, the Adam optimizer is used for the training of this model. 
The training data for this study are generated by using the isotropic Neo-Hooke model, cf. Eq.~\eqref{eq:stress_NH}, where the constants $(E, \nu) := (\SI{1}{\mega\pascal}, 0.3)$ are chosen.

\subsubsection{Interpolation behavior}
\label{subsubsec:interpolation}

In order to analyze the interpolation behavior of the three models, we investigate three different sets of uniaxial stress states: ideal data, offset data, as well as noisy data. 
\begin{figure}[b!]
	    \centering
	    \includegraphics[width=\textwidth]{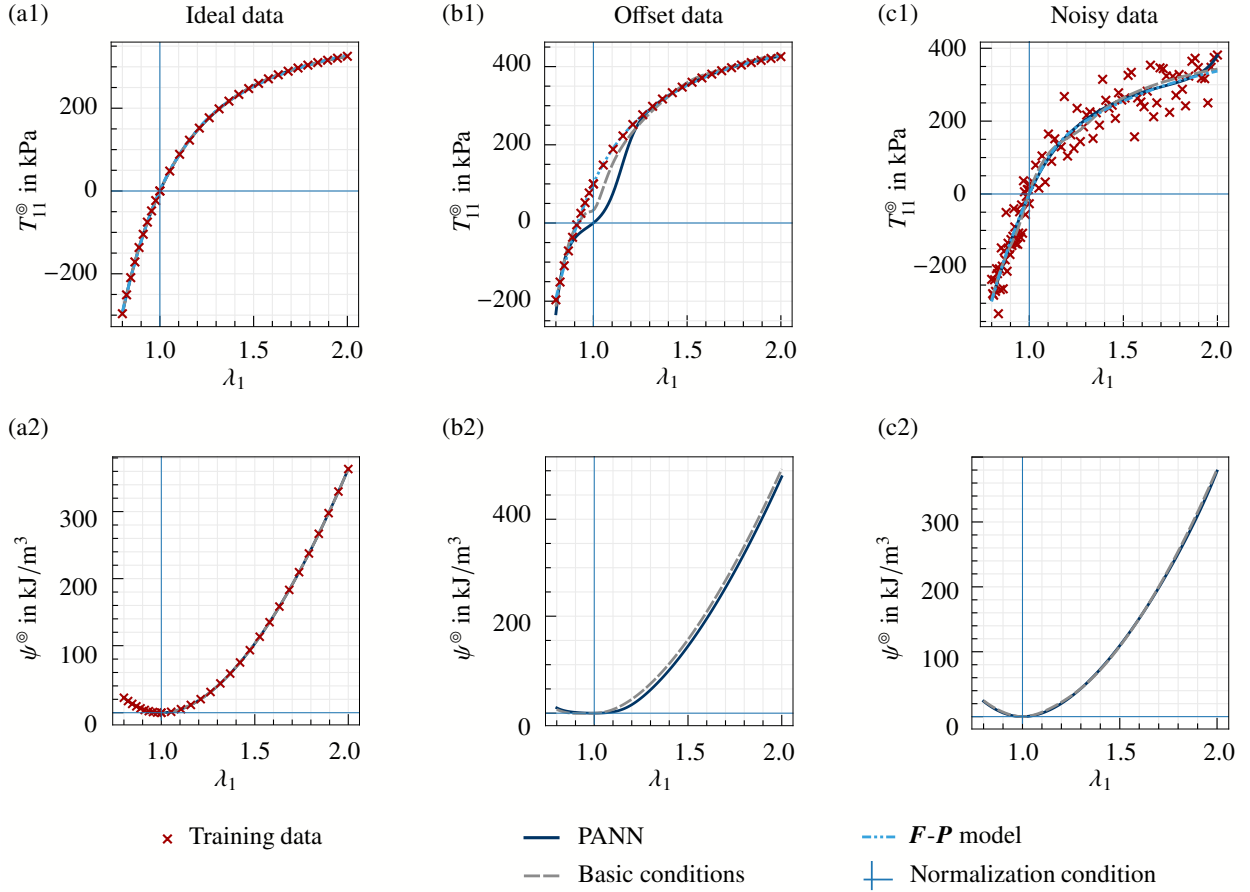}
	    \caption{Predicted stress-stretch curves $T_{11}^{\circledcirc}(\lambda_1)$ of the PANN, the NN fulfilling the basic conditions, as well as the $\te F$-$\te P$ model with one hidden layer containing $N^\text{NN} := 4$ neurons. The trained models are compared to data from a uniaxial tension/compression test in $X_1$-direction for: (a1) ideal isotropic Neo-Hooke data, (b1) data with offset, and (c1) noisy data. In (a2)--(c2), the corresponding energies of the models are shown.
	    For reasons of improved comparability, energy normalization is applied for $\psi^{\text{NN,}\circledcirc}(\ve{\mathcal I}^*)$ which, however, has no influence on the stress prediction. 
	    }
	    \label{fig:results_stress_strech}
\end{figure}

\paragraph{Ideal data} 
The three models are first trained on analytical data from a uniaxial tension and compression test in $X_1$-direction, cf. Fig.~\ref{fig:results_stress_strech}(a1). 
The principal stretch $\lambda_1 \in \R_{>0}$ in $X_1$-direction is prescribed and the principal stretch $\lambda_2 \in \R_{>0}$ has to be calculated in such a way that the stresses $T_{22}^\circledcirc, T_{33}^\circledcirc$ vanish, so that the components of the deformation tensor and the stress tensor result in
\begin{align}
\label{eq:uniaxial_ideal}
		(C_{KL})^\text{uniaxial} :=
\begin{pmatrix}
  \lambda_1^2 & 0 & 0\\ 
  0 & \lambda_2^2 & 0\\
  0 & 0 & \lambda_2^2
\end{pmatrix}\; , 
\quad
(T_{KL})^{\text{uniaxial,}\circledcirc} =
\begin{pmatrix}
  T_{11}^\circledcirc & 0 & 0\\ 
  0 & 0 & 0\\
  0 & 0 & 0
\end{pmatrix}\; .
\end{align}
Accordingly, this leads to a uniaxial stress state, which is commonly applied in experimental investigations.
Only 30 tuples $(\,{}^i \te C, {}^i \te T^\circledcirc\,)$ with $0.8\leq\lambda_1\leq 2$ are used as training data for the models.
Note that the tuple for $\lambda_1 = 1$ occurs twice in the data set, since this point is assigned to both the compression and tension ranges. 
This has no influence on the approximation behavior of the PANN model $\psi^{\text{PANN,}\circledcirc}(\ve{\mathcal I}^*)$, since this model is normalized by construction. However, for both the $\te F$-$\te P$ model and the basic conditions model $\psi^{\text{NN,}\circledcirc}(\ve{\mathcal I}^*)$, the undeformed state is weighted more heavily in the data set.

After calibration, the stress-stretch curves $T_{11}^\text{simple}(\lambda_1)$, $T_{11}^{\text{NN,}\circledcirc}(\lambda_1)$,and $T_{11}^{\text{PANN,}\circledcirc}(\lambda_1)$ of the three models are available in analytical form and are compared with the training data in Fig.~\ref{fig:results_stress_strech}(a1).
As expected, it can be seen that all three NN-based models are able to perfectly approximate the ideal dataset which is also shown by the MSEs given in Tab.~\ref{tab:MSE_interpolation}.
The basic conditions model as well as the PANN also reproduce the energy $\psi^\circledcirc(\lambda_1)$ with high accuracy although it was not trained directly, see Fig.~\ref{fig:results_stress_strech}(a2). Thus, incorporating physics does not degrade the model prediction quality.
In addition, we would like to point out that only the PANN model really fulfills the normalization condition, i.e., $T_{11}^{\text{PANN,}\circledcirc}(\lambda_1 = 1) = 0$, exactly.

\paragraph{Offset data}

The next step is to investigate the flexibility of the models for non-ideal data.
For this, the stress components of the ideal dataset are shifted according to
\begin{align}
\label{eq:uniaxial_offset}
(T_{KL})^{\text{offset,}\circledcirc} := (T_{KL})^{\text{uniaxial,}\circledcirc} \; + \;
\begin{pmatrix}
  100 & 0 & 0\\ 
  0 & 0 & 0\\
  0 & 0 & 0
\end{pmatrix}\,\si{\kilo\pascal}\,, 
\end{align}
so that the calibration dataset is no longer normalized for the undeformed state, i.e.,  $T_{11}^{\text{offset,}\circledcirc}(\lambda_1 = 1) \neq 0$. Thereby, the normalization condition included into the PANN model is of particular interest now.

As can be seen in Fig.~\ref{fig:results_stress_strech}(b1), the $\te F$-$\te P$ model $\te P^\text{simple}(\te F)$ perfectly approximates the training data again, see also the MSE given in Tab.~\ref{tab:MSE_interpolation}. However, this is only possible since this simple model does not know about the existence of a potential and is thus not really an hyperelastic one. In contrast, the basic conditions model $\psi^{\text{NN,}\circledcirc}(\ve{\mathcal I}^*)$ is not able to reproduce the training data well near $\lambda=1$, which is due to the fact that it cannot violate the material symmetry and that the stress results from the derivative of an elastic potential. However, for the undeformed state a non-zero stress is predicted, which is in contradiction to the expectation of a standard elastic model.

As can be seen in Fig.~\ref{fig:results_stress_strech}(b1), due to the model approach according to Sec.~\ref{sec:model}, the normalization condition is only exactly fulfilled by the PANN model $\psi^{\text{PANN,}\circledcirc}(\ve{\mathcal I}^*)$ after model calibration, i.e., $T_{11}^{\text{PANN,}\circledcirc}(\lambda_1 = 1) = 0$.
Although it becomes evident from the proofs in Sec.~\ref{sec:model}, it is now demonstrated that this important condition is fulfilled at all times by construction, even for the case when the training data have an offset.
Similar to the basic conditions model, due to the stress normalization and the polyconvexity, the PANN model does not approximate the data points in the neighborhood of $\lambda_1 = 1$ well -- but in a physically meaningful way -- resulting in an inflection point within the stress-stretch curve. Outside this neighborhood, the approximation of the data points is close to perfect.
Thereby, the non-convexity of $T_{11}$ in $\lambda_1$ should not be mistaken as a violation of the polyconvexity condition. The proposed NN model is polyconvex by construction, cf. Remark~\ref{rem:ICNN}, and furthermore, polyconvexity does not pose restrictions on the dependency of $\te T$ in $\te F$.

Due to the described physical restrictions, the MSEs of the basic conditions model and the PANN are several orders of magnitude larger compared to the $\te F$-$\te P$ model for the offset data, cf. Tab.~\ref{tab:MSE_interpolation}.
The elastic energies given by $\psi^{\text{NN,}\circledcirc}(\ve{\mathcal I}^*)$ and $\psi^{\text{PANN,}\circledcirc}(\ve{\mathcal I}^*)$ are shown in Fig.~\ref{fig:results_stress_strech}(b2). Thereby, no significant difference between both models occurs. Note that the training data are not included in this plot because there is no energy available for the offset dataset under consideration.

\paragraph{Noisy data}
In the last step, the approximation behavior of the three models with respect to noisy data is investigated. 
For this purpose, the stress components of a larger ideal dataset containing 100 tuples $(\,{}^i \te C, {}^i \te T^\circledcirc\,)$ are shifted according to
\begin{align}
\label{eq:uniaxial_noisy}
(T_{KL})^{\text{noisy,}\circledcirc} := (T_{KL})^{\text{uniaxial,}\circledcirc} \; + \;
\begin{pmatrix}
  \xi & 0 & 0\\ 
  0 & 0 & 0\\
  0 & 0 & 0
\end{pmatrix} \quad \text{with} \quad \xi \sim \mathcal N(\mu, \sigma^2)\; ,
\end{align}
where $\xi$ describes a normally distributed noise with mean value $\mu = \SI{0}{\kilo\pascal}$ and standard deviation $\sigma = \SI{50}{\kilo\pascal}$.

Although the training data are overlaid with a strong Gaussian noise, as it can be seen in Fig.~\ref{fig:results_stress_strech}(c1), all three calibrated models give a result that appears to be physically reasonable. 
To be more precise, the predicted stress-stretch curves are monotonously increasing and the potentials $\psi^{\text{PANN,}\circledcirc}(\lambda_1)$ and $\psi^{\text{NN,}\circledcirc}(\lambda_1)$ shown in Fig.~\ref{fig:results_stress_strech}(c2) are convex in $\lambda_1$, which is both to be expected for uniaxial stress, they are smooth, and do not show any markedly oscillations.
It can be seen that the stress prediction is approximately zero for the undeformed state in this case, see Fig.~\ref{fig:results_stress_strech}(c1).
However, only the PANN model satisfies all relevant conditions in an exact manner, which also applies to stress-strain states not included in the training. The advantage of this property will be revealed in the next study. The $\te F$-$\te P$ model even violates the balance of angular momentum, since $\te T^{\text{simple}}$ and $\te \sigma^{\text{simple}}$ are not symmetrical.

The MSEs for all three models are again given in Tab.~\ref{tab:MSE_interpolation}. Due to the imperfect data, the values are increased by several orders of magnitude compared to the ideal case.

\begin{table}[th!]
\renewcommand{\arraystretch}{1.5}
\centering
\caption{MSE according to Eq.~\eqref{eq:mse_calibration} achieved with the $\te F$-$\te P$ model, the NN fulfilling the basic conditions and the PANN for training with different uniaxial stress data. The respective stress predictions are shown in Fig.~\ref{fig:results_stress_strech}.}
\begin{small}
\begin{tabular}{c c c c}
    Model & Ideal data & Data with offset & Noisy data \\
    \hline
    $\te F$-$\te P$ model & \SI{3.14e0}{\kilo\pascal\squared} & \SI{2.87e0}{\kilo\pascal\squared} & \SI{2.02e3}{\kilo\pascal\squared} \\
    Basic conditions NN & \SI{1.74e-4}{\kilo\pascal\squared} & \SI{1.22e3}{\kilo\pascal\squared} & \SI{2.03e3}{\kilo\pascal\squared} \\
    PANN model & \SI{5.92e-5}{\kilo\pascal\squared} & \SI{2.77e3}{\kilo\pascal\squared} & \SI{2.02e3}{\kilo\pascal\squared}
\end{tabular}
\end{small}
\label{tab:MSE_interpolation}
\end{table}

\subsubsection{Extrapolation behavior}
\label{subsubsec:extrapolation}

Now, the extrapolation behavior of the compared models is analyzed by considering three load cases: uniaxial tension/compression, biaxial tension/compression, as well as simple shear. For the training of $\te P^{\text{simple}}(\te F)$, $\psi^{\text{NN,}\circledcirc}(\ve{\mathcal I}^*)$, and $\psi^{\text{PANN,}\circledcirc}(\ve{\mathcal I}^*)$ only uniaxial stress states and the corresponding deformations within the narrow range $0.8\le\lambda_1\le 1.1$ have been used here, cf. Eq.~\eqref{eq:uniaxial_ideal}, which are stored in the dataset $\mathcal D^{\text{uniaxial,}\circledcirc}$ with $\vert\mathcal D^{\text{uniaxial,}\circledcirc}\vert = 15$.
Again, note that the tuple for $\lambda_1 = 1$ occurs twice in the data set, since this point is assigned to both the compression and tension ranges. 
Given that the PANN model $\psi^{\text{PANN,}\circledcirc}(\ve{\mathcal I}^*)$ is normalized by construction, this does not affect the approximation behavior.
In contrast, for both the $\te F$-$\te P$ model and the basic conditions model $\psi^{\text{NN,}\circledcirc}(\ve{\mathcal I}^*)$, the undeformed state is weighted more heavily in the data set.

\begin{figure}[bh!]
	    \centering
	    \includegraphics[width=\textwidth]{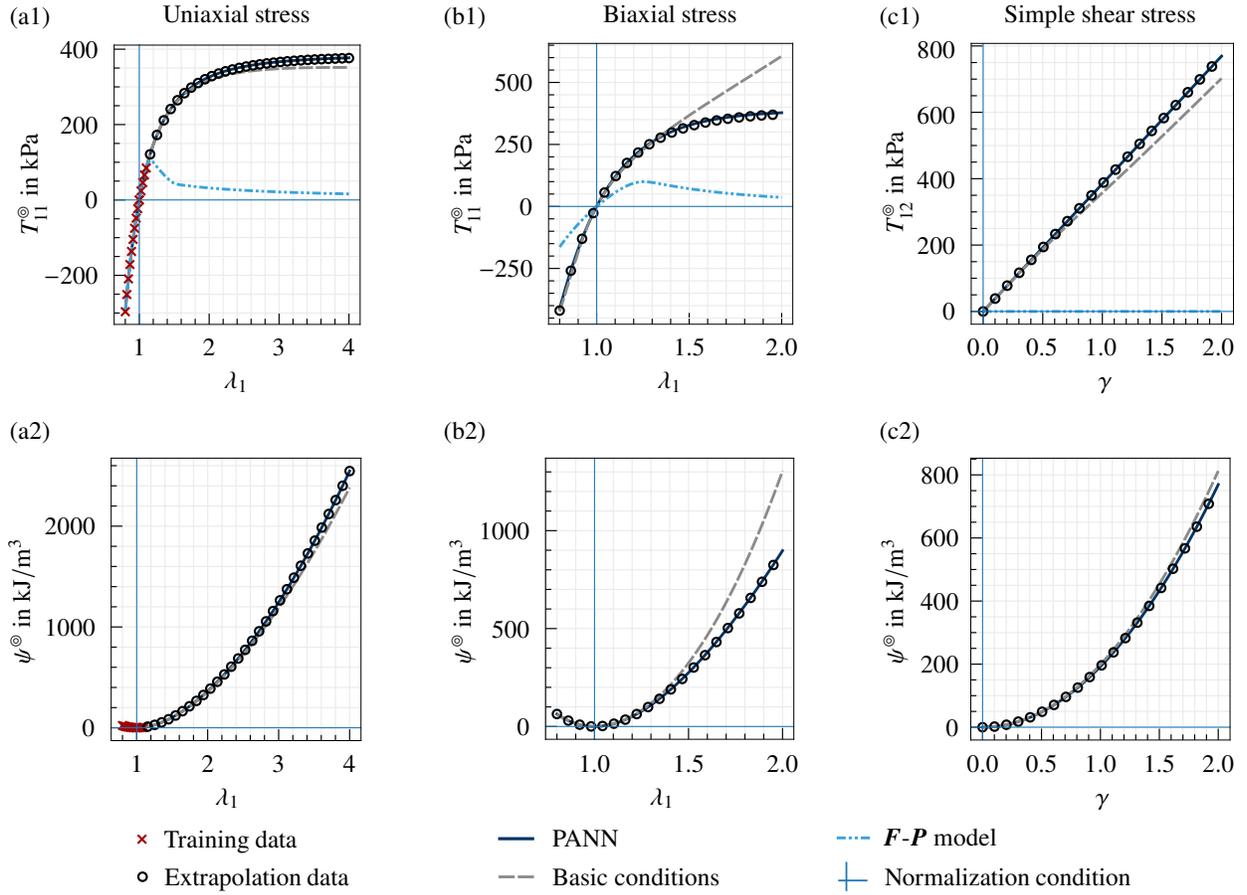}
	    \caption{Predicted stress-stretch curves $T_{KL}^{\circledcirc}(\lambda_1)$ and $T_{KL}^{\circledcirc}(\gamma)$ of the PANN, the NN fulfilling the basic conditions, as well as the $\te F$-$\te P$ model with one hidden layer containing $N^\text{NN} := 4$ neurons. Shown is the extrapolation of the models trained with uniaxial stress data stored in $\mathcal D^{\text{uniaxial,}\circledcirc}$ from an isotropic Neo-Hooke model for: (a1) uniaxial tension/compression test, (b1) biaxial tension/compression test, and (c1) a  simple shear test. The corresponding energies are given in (a2)--(c2).
	    }
	    \label{fig:results_extrapolation}
	\end{figure}
\begin{figure}[h!]
	    \centering
	    \includegraphics[width=\textwidth]{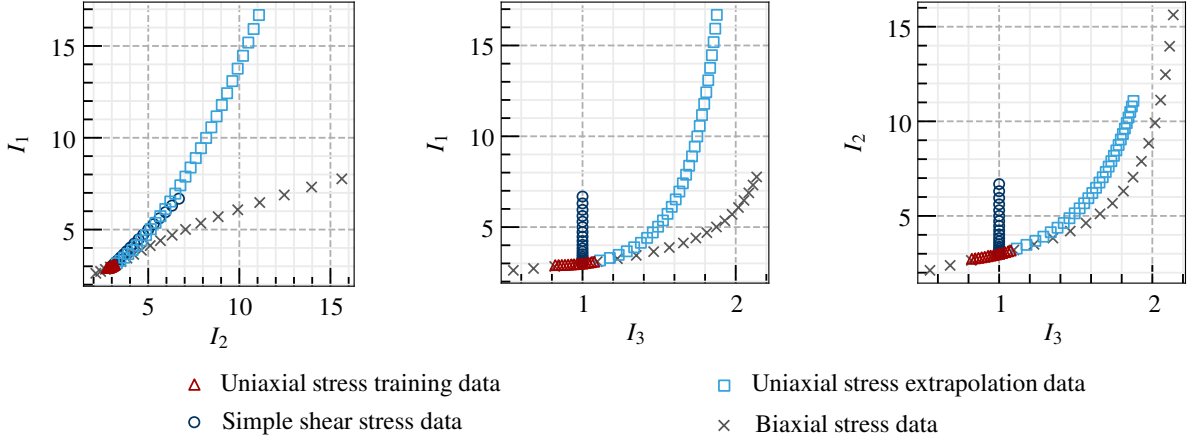}
	    \caption{Location of the deformation states within the isotropic invariant space for uniaxial tension/compression test, biaxial tension/compression test, and simple shear test. Shown are the sectional planes $I_1$-$I_2$, $I_1$-$I_3$, and $I_2$-$I_3$.
	    }
	    \label{fig:results_extrapolation_invariants}
	\end{figure}

\paragraph{Uniaxial stress states}	
As one can see in Fig.~\ref{fig:results_extrapolation}(a1), all models again perfectly approximate the training data which is also evident from the MSEs given in Tab.~\ref{tab:MSE_extrapolation}. However, the $\te F$-$\te P$ model fails immediately when it has to extrapolate. The model $\psi^{\text{NN,}\circledcirc}(\ve{\mathcal I}^*)$, in contrast, is very good at extrapolating up to a stretch of $\lambda_1=4$, which is really impressing from the author's point of view.
Likewise, the related energy shown in Fig.~\ref{fig:results_extrapolation}(a2) is very well reproduced.
Thus, the advantage of invariant-based approaches  \cite{Fuhg2022b,kalina2022b,klein2021,klein2022a,tac2022,Linden2021,Linka2020,linka2023} which approximate the energy and not directly the stress is particularly evident here.
Even more impressive is the result for the PANN model, which is able to reproduce the data almost perfectly despite extrapolation. This is also evident from the MSE for the extrapolated data, which is two orders of magnitude lower for the PANN model than for the basic conditions model, cf. Tab.~\ref{tab:MSE_extrapolation}.
The observed improvement results from the insertion of the additional physical principles, namely polyconvexity, growth condition, as well as energy and stress normalization, into the PANN model.

\paragraph{Biaxial stress states and simple shear}	
Now, as a next step, we want to evaluate how the models perform if they have to extrapolate for completely unknown load cases, i.e., biaxial stress states
\begin{align}
\label{eq:biaxial_ideal}
		(C_{KL})^\text{biaxial} :=
\begin{pmatrix}
  \lambda_1^2 & 0 & 0\\ 
  0 & \lambda_1^2 & 0\\
  0 & 0 & \lambda_2^2
\end{pmatrix}\; , 
\quad
(T_{KL})^{\text{biaxial,}\circledcirc} =
\begin{pmatrix}
  T_{11}^\circledcirc & 0 & 0\\ 
  0 & T_{11}^\circledcirc & 0\\
  0 & 0 & 0
\end{pmatrix}\; 
\end{align}
and simple shear
\begin{align}
\label{eq:shear_ideal}
		(C_{KL})^\text{shear} :=
\begin{pmatrix}
  1 & \gamma & 0\\ 
  \gamma & \gamma^2+1 & 0\\
  0 & 0 & 1
\end{pmatrix}\; 
\end{align}
with $0 \leq \gamma\leq 2$ denoting the shearing. The results for the stress predictions are given in Fig.~\ref{fig:results_extrapolation}(b1) and (c1). Again, the simple approach $\te P^{\text{simple}}(\te F$) completely fails for the unseen data. In the simple shear load case, even a completely implausible shear stress $T^{\text{simple}}_{12}\approx 0$ is predicted, since the model has learned in training that the diagonal elements of the stress tensor disappear. In contrast, the basic conditions model $\psi^{\text{NN,}\circledcirc}(\ve{\mathcal I}^*)$ is still quite good at physically plausible extrapolation, but shows noticeable deviations for $\lambda_1>1.4$ within the biaxial stress loading. A similar result is obtained for simple shear, where noticeable deviations occur starting at $\gamma\approx 0.8$. The same holds for the approximation of the energy which is shown in Fig.~\ref{fig:results_extrapolation}(b2) and (c2).
Surprisingly, the PANN model is able to predict the data almost perfectly for biaxial and simple shear, although full extrapolation is required here. 
The significantly improved extrapolation capability of the PANN model $\psi^{\text{PANN,}\circledcirc}(\ve{\mathcal I}^*)$ compared to $\te P^{\text{simple}}(\te F)$ and $\psi^{\text{NN,}\circledcirc}(\ve{\mathcal I}^*)$ becomes also evident from the MSEs given in Tab.~\ref{tab:MSE_extrapolation}.
To illustrate into which ranges the models must extrapolate for the considered test cases, the deformation states examined are shown within the invariant space in Fig.~\ref{fig:results_extrapolation_invariants}.

Thus, summarizing the findings of the presented study, the adaption of NN-based models in such a way that they fulfill physical conditions for arbitrary loadings does not necessarily improve the approximation of training data, at least for the simple load cases considered, but it does allow for a significant improvement in the extrapolation capability.

\begin{table}[th!]
\renewcommand{\arraystretch}{1.5}
\centering
\caption{MSE according to Eq.~\eqref{eq:mse_calibration} achieved with the $\te F$-$\te P$ model, the NN fulfilling the basic conditions, and the PANN model for training with uniaxial tension/compression ($0.8\le\lambda_1\le 1.1$) as well as extrapolation to unknown states.}
\begin{small}
\begin{tabular}{c c c c c}
    Model & Training uniaxial stress & Extrapolation uniaxial stress &  Biaxial stress & Simple shear stress \\
    \hline
    $\te F$-$\te P$ model & \SI{1.95e-1}{\kilo\pascal\squared} & \SI{9.43e4}{\kilo\pascal\squared} & \SI{1.35e5}{\kilo\pascal\squared} & \SI{8.06e5}{\kilo\pascal\squared} \\
    Basic conditions NN & \SI{1.30e-4}{\kilo\pascal\squared} & \SI{1.26e5}{\kilo\pascal\squared} & \SI{6.24e5}{\kilo\pascal\squared} & \SI{5.65e4}{\kilo\pascal\squared}\\
    PANN model & \SI{3.91e-5}{\kilo\pascal\squared} & \SI{6.21e2}{\kilo\pascal\squared} & \SI{4.11e3}{\kilo\pascal\squared} & \SI{1.58e-5}{\kilo\pascal\squared}
\end{tabular}
\end{small}
\label{tab:MSE_extrapolation}
\end{table}

\subsection{Complex multiaxial stress-strain states}
\label{subsec:multiaxial_deformation}

Now, as we have shown the advantage of \emph{invariant-energy-based NN approaches}, we want to analyze only this model class in the following.
Thereby, it is of interest whether the exact fulfillment of the physical principles introduced in Sec.~\ref{subsec:requirements} could be too restrictive, so that the model is not flexible enough to approximate complex multiaxial stress-strain states given by a non-linear material behavior sufficiently well.
Thereby, the four different architectures $\psi^{\text{(i),}\square}(\ve{\mathcal I}^*)$, $\psi^{\text{(ii),}\square}(\ve{\mathcal I}^*)$, $\psi^{\text{(iii),}\square}(\ve{\mathcal I}^*)$, as well as $\psi^{\text{(iv),}\square}(\ve{\mathcal I}^*)$ satisfying 
\begin{enumerate}
	\item[(i)] the basic conditions, i.e., thermodynamic consistency, symmetric $\te \sigma$, objectivity and material symmetry,
	\item[(ii)] the conditions of (i) + polyconvexity,
	\item[(iii)] the conditions of (ii) + growth condition, as well as
	\item[(iv)] the conditions of (iii) + energy and stress normalization
\end{enumerate}
are compared to each other for isotropic as well as transversely isotropic behavior, respectively. Note that the first and the fourth model are equal to $\psi^{\text{NN,}\square}(\ve{\mathcal I}^*)$ and $\psi^{\text{PANN,}\square}(\ve{\mathcal I}^*)$.

\subsubsection{Generation of training data}
\label{subsubsec:data_generation}

\begin{figure}[th]
	    \centering
	    \includegraphics[width=0.6\textwidth]{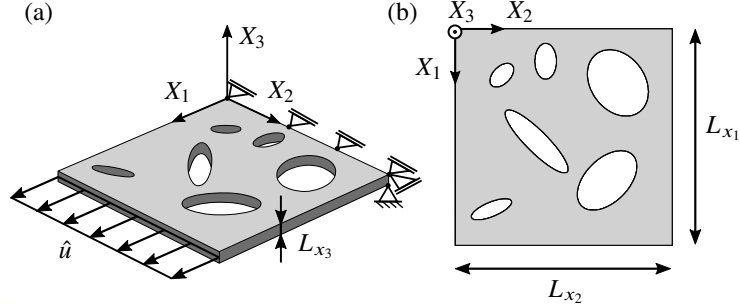}
	    \caption{Uniaxial tensile test for data generation: (a) three-dimensional, inhomogeneous specimen and applied boundary conditions with prescribed displacement $\hat{u}\, \ve e_1$ as well as (b) top view of the specimen's geometry with $L_{x_1} \times L_{x_2} \times L_{x_3} = (100 \times 100 \times 5) \, \si{\milli\metre}$.}
	    \label{fig:data_generation}
	\end{figure}

In a first step, the data basis for the training of the NNs has to be acquired. 
In the absence of real experiments, it is generated numerically here.
To this end, a uniaxial tensile test is applied on a virtual sample in an FE simulation, see Fig.~\ref{fig:data_generation}(a)~and~(b).
By applying a uniaxial tensile test on this structure, the holes in the structure will lead to fairly inhomogeneous deformation gradients at each quadrature point of the finite element mesh. The aim of this simulation is to generate such general deformation gradients, which are then used to generate data with the analytical models introduced in Sec.~\ref{sec:spec_aniso}.
Here, the sample's geometric dimensions are specified by $L_{x_1} \times L_{x_2} \times L_{x_3} = (100 \times 100 \times 5) \, \si{\milli\metre}$.
Within the uniaxial tensile test, the displacement boundary condition with a maximum value $\hat u \leq  \SI{40}{\milli\meter}$ is prescribed, i.e., $u_1 = \hat u$ holds in the plane $X_1 = L_{x_1}$ and is linearly increased in each increment of the FE simulation.
In contrast, $u_1 = 0$ holds in the plane $X_1 = 0$.
At the point $(0, L_{x_2}, 0)$ the displacement $u_2 = 0$ is additionally prescribed and due to the fixed bearing at the point $(0, L_{x_2}, -L_{x_3})$ the displacement is blocked, i.e., $\ve u = \ve 0$ holds in order to prevent rigid body motion.
The geometry generation and the meshing was realized using the tool \textit{Gmsh}~\cite{Geuzaine2009}.

Within this virtual experiment, the required dataset $\mathcal D^\circledcirc$ consisting of the tuples $\mathcal D_i^\circledcirc := ({}^i \te C, {}^i \te T^\circledcirc)$ is collected at the quadrature points of the finite elements within 30 increments.
In order to demonstrate the ability of the proposed NN-based method, the nonlinear stress-strain relation~\eqref{eq:stress_NH} is chosen for the isotropic constitutive behavior of the sample's material. 
Here the material parameters for the analytical model are chosen according to Tab.~\ref{tab:material_parameters}.

Following the work of Kalina~et~al.~\cite{kalina2022a} and prescribing  a relative tolerance $\eta := \SI{1}{\percent}$, the dataset was then filtered w.r.t. the invariants $(\, {}^i I_1,{}^i I_2, {}^i I_3\,) \in \R^3$ for the corresponding deformation state ${}^i \te C$ to obtain a reduced dataset $\mathcal D^{\text{red},\circledcirc}$ with $\vert\mathcal D^{\text{red},\circledcirc}\vert = 963$ for the calibration process of the PANN. 
This procedure took advantage of the fact that the introduced model lives in the space of invariants rather than in the space of deformations.

Finally, different deformation states are available, which are stored for the isotropic case with the corresponding stresses of the Neo-Hooke model in the dataset $\mathcal D^{\text{red},\circledcirc}$.
For the transversely isotropic case with the preferred direction parallel to the $X_1$-direction, the same deformation states are chosen as data basis and stored with the corresponding stresses of Schröder's model~\eqref{eq:stress_ti} in the dataset $\mathcal D^{\text{red}, \parallel}$ with $\vert\mathcal D^{\text{red},\parallel}\vert = 963$ as well. The chosen parameters of the analytical transversely isotropic model are also given in Tab.~\ref{tab:material_parameters}.

\begin{table}[h]
\renewcommand{\arraystretch}{1.5}
\centering
\caption{Material parameters of isotropic Neo-Hooke and transversely isotropic analytical model proposed by Schr\"oder et al.~\cite{Schroeder2008}. The models are given in Eqs.~\eqref{eq:stress_NH} and \eqref{eq:stress_ti}, respectively.}
\begin{small}
\begin{tabular}{ c  c || c  c  c  c  c  c  c } 
  $E$ &  $\nu$ &  $\beta$ &   $\alpha_1$ &   $\alpha_2$ &   $\delta_1$ &   $\delta_2$ &   $\alpha_4$ &   $\eta_1$
 \\[0.5ex]
 \hline
$\SI{e3}{\kilo\pascal}$ & $0.3$ & $2$ & $\SI{8}{\kilo\pascal}$ & $\SI{0}{\kilo\pascal}$ & $\SI{10}{\kilo\pascal}$ & $\SI{56}{\kilo\pascal}$ & $2$ & $\SI{10}{\kilo\pascal}$
\end{tabular}
\end{small}
\label{tab:material_parameters}
\end{table}

\subsubsection{Overall prediction quality}
\label{subsubsec:prediction_quality}
\begin{figure}[t]
	    \centering
	    \includegraphics[width=\textwidth]{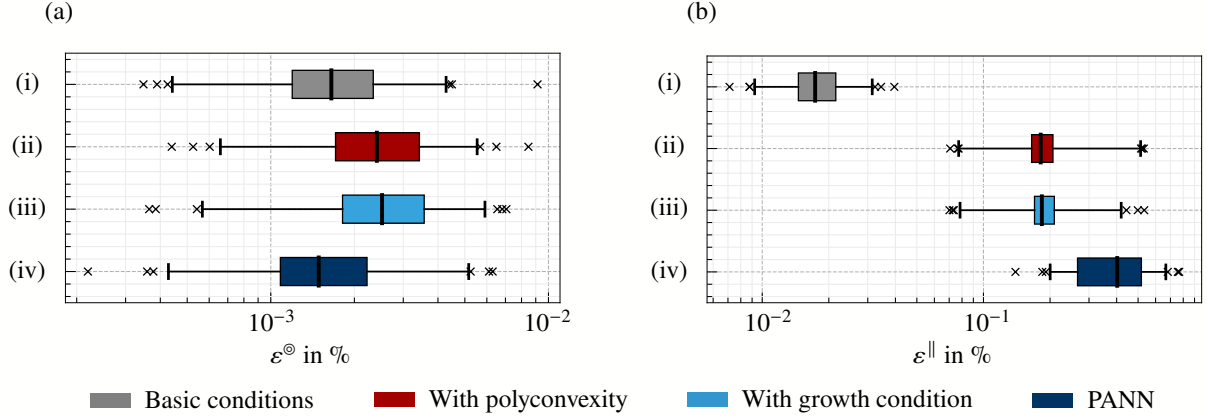}
	    \caption{Boxplots with the median $\varepsilon^{\text{med,}\square}$, 25th and 75th as well as the 1st and 99th percentile of relative error measure $\varepsilon^\square$ according to Eq.~\eqref{eq:error_relative}  for the reduced dataset $\mathcal D^{\text{red},\square}$ and the (a) isotropic NNs or (b) transversely isotropic NNs with (i) basic conditions, (ii) polyconvexity, (iii) growth condition and polyconvexity as well as (iv) the PANN satisfying all conditions including normalization.
	    }
	    \label{fig:error_ANN_conditions}
	\end{figure}

First, for the reduced dataset $\mathcal D^{\text{red},\square}$, we will compare the overall prediction quality of the trained NNs (i)--(iv) satisfying different physical constraints, introduced in Sec.~\ref{subsec:requirements}, for the isotropic as well as the transversely isotropic case. Thereby, for the transversely isotropic case $\beta = 2$ is chosen for the structural tensor $\te G$ given in Eq.~\eqref{eq:structural_tensor}.
In the following examples, the network architecture is set to one hidden layer with $N^\text{NN} := 8$ neurons.
The NN-based models are trained w.r.t. the dataset $\mathcal D^{\text{red},\square}$, where a random division into calibration (70\,\%) and test (30\,\%) data is made once.
Within the training process, the weights and bias values  $W_\alpha, b_\alpha, w_{\alpha\beta}$ and $w_{\alpha\gamma}^*$ are then determined according to Sec.~\ref{subsec:model_calibration}. 
Within one training run, the respective NN is trained 30 times, and the parameters of the best achieved training state with the lowest MSE~\eqref{eq:mse_calibration}, see Sec.~\ref{subsec:model_calibration}, are stored at the end \cite{kalina2022a}.\footnote{Due to local minima within the loss function, the optimization procedure which is applied here depends on the starting values of the weights and biases. Thus, the network is trained several times to overcome this, cf. Kalina~et~al.~\cite{kalina2022a,kalina2022b}.}
In order to evaluate the approximation behavior of the trained NNs, we compute the relative error measure
\begin{align}
	\label{eq:error_relative}
    \varepsilon^\square := \frac{\max\limits_{i \in \N_{\leq \vert\mathcal D^\square\vert}} \Big\Vert  {}^i\te T^\square - {}^i\te T^{\text{model,}\square}\Big\Vert}{\max\limits_{j \in \N_{\leq \vert\mathcal D^\square\vert}} \|{}^j\te T^\square\|}
\end{align}
for the Frobenius norm $\|\te T\|$ of the second Piola-Kirchhoff stress. To exclude random effects, a statistical study is performed, i.e., a total of 300 training runs has been performed for each model. 
Since no uniform distribution w.r.t. the error measure $\varepsilon^\square$ can be seen from the results, no underlying distribution is assumed here, see the histogram plots given in App.~\ref{sec:stochastic}.
Thus, median $\varepsilon^{\text{med,}\square}$ and quantiles are used to compare the interpolation quality of the models with each other. The results of the described study are shown in Fig.~\ref{fig:error_ANN_conditions}.

\paragraph{Isotropic model}
Regarding the NNs' stress approximation for the isotropic case, an extremely good prediction quality with a median of the errors $\varepsilon^{\text{med,}\circledcirc} < \SI{0.003}{\percent}$ is achieved for all models, cf. Fig.~\ref{fig:error_ANN_conditions}(a). %
As one can see, the approximation quality of $\psi^{\text{(ii),}\circledcirc}(\ve{\mathcal I}^*)$ accounting for the polyconvexity worsens in comparison to the architecture $\psi^{\text{(i),}\circledcirc}(\ve{\mathcal I}^*)$ which only fulfills the basic conditions.
Thus, the limitation to positive weights reduces the approximation quality of the NN in the statistical sense. However, as already mentioned, the errors are still extremely low.
If the growth condition is further added, the error of this NN denoted as $\psi^{\text{(iii),}\circledcirc}(\ve{\mathcal I}^*)$, which fulfills basic conditions + polyconvexity + growth condition, stays in a similar range. 
Surprisingly, when the normalization conditions are finally added, the resulting distribution achieved with $\psi^{\text{(iv),}\circledcirc}(\ve{\mathcal I}^*)$ is very similar to the first model, i.e., the basic conditions model.
Thus, summarizing the study carried out, adding all common physical principles of hyperelasticity to the NN-based model does not lead to a deterioration in the approximation for the isotropic case.
This is although conditions such as positive weights to account for polyconvexity are reducing the variability of the NN.

Finally, regarding the non-negativity condition on the elastic energy, a numerical test has been applied. In order to scan only physically admissible deformation states, the principal stretch $\lambda \in \R_{>0}$ is varied and $\te C^\text{diag}(\lambda):=\lambda^2 \te 1$ is computed as a spherical tensor, which is sufficient for isotropy according to  Theorem~\ref{theorem:isotropic_PANN} of App.~\ref{app:non_neg}. The invariants are calculated and the respective energy is determined. Within a range of $1/10\leq\lambda\leq 10$, only positive energies have been numerically detected for the trained isotropic PANN with relative error $\varepsilon^\circledcirc$, which is closest to the median $\varepsilon^{\text{med,}\circledcirc}$, exemplarily.

\paragraph{Transversely isotropic model}
Regarding the NNs' stress approximation for the transversely isotropic scenario given in Fig.~\ref{fig:error_ANN_conditions}(b), compared to isotropy, the error is now an order of magnitude higher even for the best model $\psi^{\text{(i),}\parallel}(\ve{\mathcal I}^*)$. This can be explained by the increased complexity of the reference model \eqref{eq:energy_ti}.
However, with a median of the errors $\varepsilon^{\text{med,}\parallel} < \SI{0.4}{\percent}$, a high level of prediction quality is still attained for all models.
Now the four models (i)--(iv) are compared. Applying the model $\psi^{\text{(ii),}\parallel}(\ve{\mathcal I}^*)$, which takes polyconvexity into account, leads to errors that are increased by an order of magnitude compared to the basic conditions architecture $\psi^{\text{(i),}\parallel}(\ve{\mathcal I}^*)$.
Thus, the limitation to positive weights significantly reduces the approximation quality of the NN. If the growth condition is further added, the error of this NN, denoted as $\psi^{\text{(iii),}\parallel}(\ve{\mathcal I}^*)$, stays in a similar range. 
Finally, when the normalization conditions are added, the resulting distribution of erros achieved with $\psi^{\text{(iv),}\parallel}(\ve{\mathcal I}^*)$ shifts again to slightly larger values.

We would like to emphasize that the models are based on only one hidden layer with $N^\text{NN} = 8$ neurons. The approximation quality of the NN-based models could easily be increased by adapting the network architecture, i.e., using several hidden layers or more neurons, in order to be able to represent the complex material behavior sufficiently well.
Summarizing, adding all common physical principles of hyperelasticity to the NN-based model leads to a deterioration of the prediction quality of approximately one order of magnitude for the transversely isotropic case. Since the errors are nevertheless very small, the PANN model should be chosen anyway, especially with regard to the very good extrapolation capability.

Regarding the non-negativity condition on the elastic energy, a numerical test has been applied for transverse isotropy, too. Again, to scan only physically admissible deformation states, the principal stretches $\lambda_1,\lambda_2,\lambda_3 \in \R_{>0}$ are varried and the diagonal tensor $\te C^\text{diag}(\lambda_1,\lambda_2,\lambda_3) = \lambda_1^2 \ve e_1 \otimes \ve e_1 + \lambda_2^2 \ve e_2 \otimes \ve e_2 + \lambda_3^2 \ve e_3 \otimes \ve e_3$ is computed. Here, in addition, rotations of these states perpendicular to the preferred direction $X_1$ have to be considered. Thus, we end up with a five parameter space for the deformation to be sampled:
\begin{align}
    \te C^\parallel(\lambda_1,\lambda_2,\lambda_3,\varphi_2,\varphi_3) =
    \te R(\varphi_2,\varphi_3) \cdot \te C^\text{diag}(\lambda_1,\lambda_2,\lambda_3)
    \cdot \te R^T(\varphi_2,\varphi_3)
    \; .
    \label{eq:C_transverse}
\end{align}
In the equation above, $\te R(\varphi_2,\varphi_3) = \te R_{x_2}(\varphi_2) \cdot \te R_{x_3}(\varphi_3) \in \SO(3)$ is a rotation tensor with $\te R_{x_2}(\varphi_2)$ and $\te R_{x_2}(\varphi_3)$ denoting rotations around the $X_2$- and $X_3$-axis, respectively.
With Eq.~\eqref{eq:C_transverse} and the structural tensor given in Eq.~\eqref{eq:structural_tensor} invariants are calculated and the respective energy is determined. Within a range of $1/10\le\lambda_\alpha\le 10$, $0\le\varphi_\beta\le \pi/2$ with $\alpha\in\N_{\leq 3}, \beta \in\{2, 3\}$, only positive energies have been numerically detected for both the model~\eqref{eq:energy_ti} and the trained transversely isotropic PANN with relative error $\varepsilon^\parallel$, which is closest to the median $\varepsilon^{\text{med,}\parallel}$, exemplarily.
\subsubsection{Application of the calibrated PANN within an FE simulation}
\label{subsubsec:fe_application}
\begin{figure}[th!]
	    \centering
	    \includegraphics[width=\textwidth]{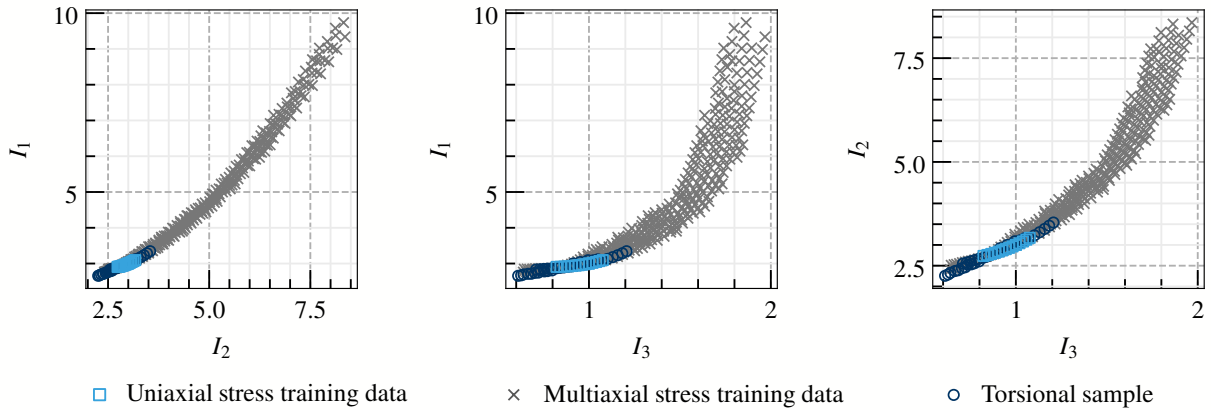}
	    \caption{Location of the deformation states within the invariant space for uniaxial stress-strain data $\mathcal D^{\text{uniaxial,}\circledcirc}$, multiaxial stress-strain data $\mathcal D^{\text{red,}\circledcirc}$ and the torsional sample, which is loaded by specifying a distortion of $\hat \phi = \SI{45}{\degree}$, cf.~Fig.~\ref{fig:results_torsion_ellipse_training}(a). 
	    Shown are the sectional planes $I_1$-$I_2$, $I_1$-$I_3$, and $I_2$-$I_3$.
	    }
	    \label{fig:results_FEM_invariants}
	\end{figure}
\begin{figure}[b!]
	    \centering
	    \includegraphics[width=\textwidth]{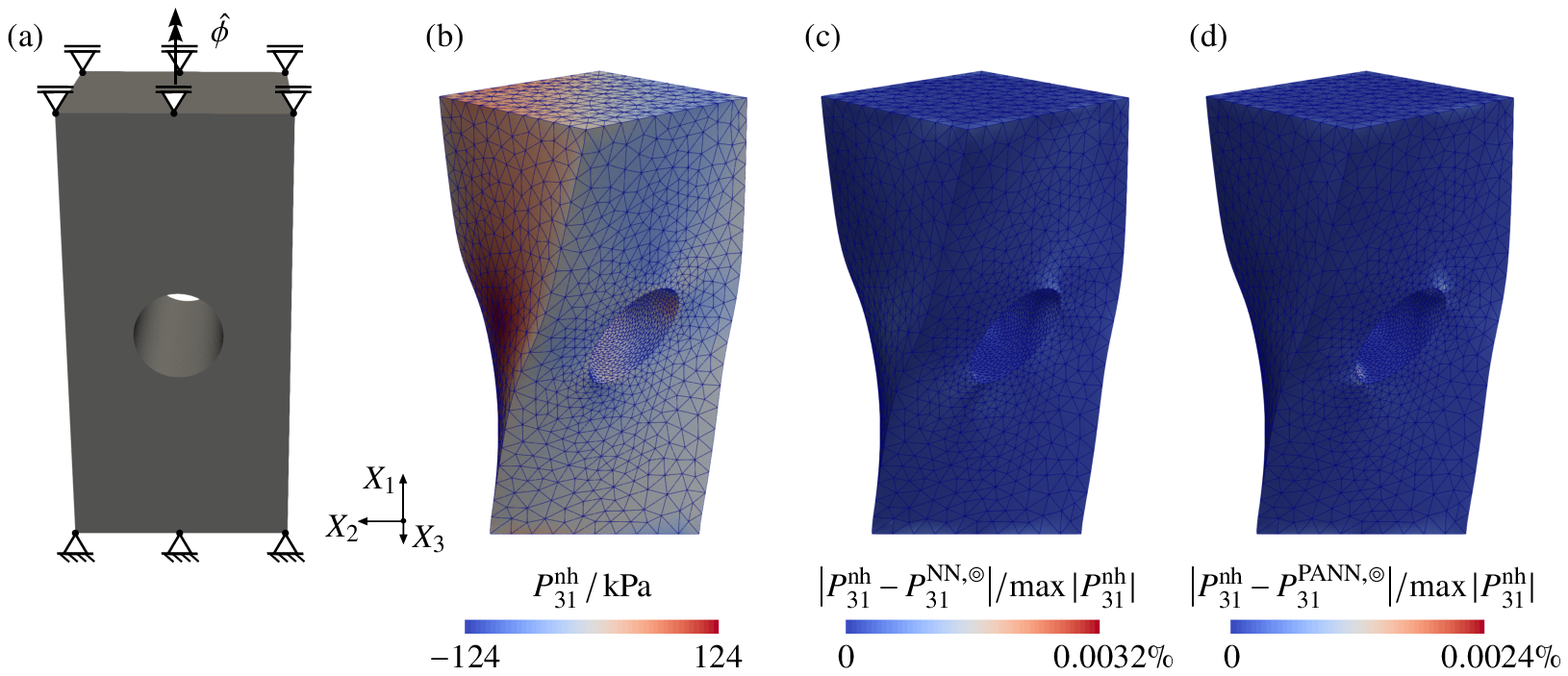}
	    \caption{FE simulation of torsional sample without warping: (a) applied boundary conditions, (b) macroscopic stress field $ P_{31}^\text{nh}$ on the deformed configuration $\mathcal B$ by specifying $\hat \phi = \SI{45}{\degree}$ as angle of twist, and (c) relative error of the PANN-stress field $P_{31}^{\text{PANN},\circledcirc}$ as well as (d) relative error of the stress field $P_{31}^{\text{NN},\circledcirc}$ of basic conditions model w.r.t. $P_{31}^\text{nh}$. The NNs were trained with multiaxial stress-strain data $\mathcal D^{\text{red},\circledcirc}$, $\vert\mathcal D^{\text{red},\circledcirc}\vert = 963$, and $N^\text{NN} = 8$ neurons in only one hidden-layer were implemented as constitutive equations each.}
	    \label{fig:results_torsion_ellipse_training}
	\end{figure}
\begin{figure}[b!]
	    \centering
	    \includegraphics[width=\textwidth]{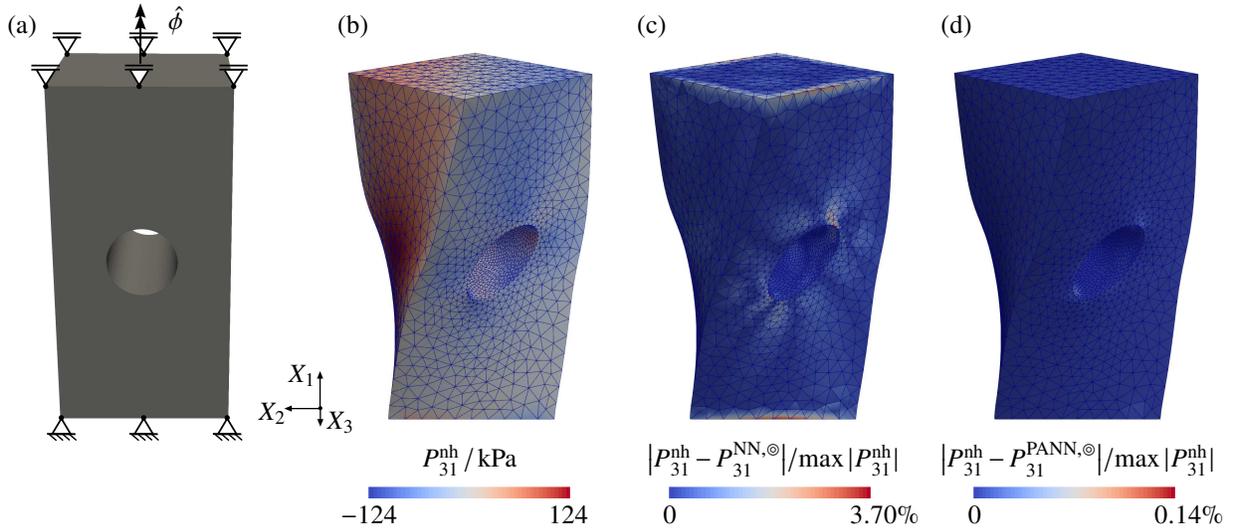}
	    \caption{FE simulation of torsional sample without warping: (a) applied boundary conditions, (b) macroscopic stress field $ P_{31}^\text{nh}$ on the deformed configuration $\mathcal B$ by specifying $\hat \phi = \SI{45}{\degree}$ as angle of twist, and (c) relative error of the PANN-stress field $P_{31}^{\text{PANN},\circledcirc}$ as well as (d) relative error of the stress field $P_{31}^{\text{NN},\circledcirc}$ of basic conditions model w.r.t. $P_{31}^\text{nh}$. The NNs were trained with only uniaxial stress-strain data $\mathcal D^{\text{uniaxial},\circledcirc}$, $\vert\mathcal D^{\text{uniaxial},\circledcirc}\vert = 15$, and $N^\text{NN} = 4$ neurons in only one hidden-layer were implemented as constitutive equations each.}
	    \label{fig:results_torsion_uniaxial_training}
	\end{figure}
In order to prove the suitability of invariant-based NNs for the numerical simulation of complex shaped samples and components, a comparison to reference results generated with the isotropic Neo-Hooke model~\eqref{eq:stress_NH}, which has been used for the calibration, is shown in the following. Thereby, the model fulfilling the basic conditions and the PANN are each analyzed in two different scenarios: once trained with multiaxial stress-strain states $\mathcal D^{\text{red},\circledcirc}$ and once with only uniaxial stress-strain states $\mathcal D^{\text{uniaxial,}\circledcirc}$, cf. Sects.~\ref{subsubsec:extrapolation} and \ref{subsubsec:prediction_quality}. In the first case, one of the 300 trained models is selected that is closest to the median $\varepsilon^{\text{med,}\circledcirc}$.

As a test example, a three-dimensional problem is considered: the torsion of a prismatic sample, cf. the sketch given in Fig.~\ref{fig:results_torsion_ellipse_training}(a).
Thus, regarding the nonlinear character of the reference constitutive law $\psi^\text{nh}(I_1,I_3)$, this example is a good possibility to validate and explore potential limits of both approaches.
In the validation setup, the three-dimensional torsion sample with one circular hole is loaded by specifying $\hat \phi = \SI{45}{\degree}$ as angle of twist in the plane $\max_{\ve X \in \mathcal B_0} X_1$. Additionally, inside this plane the displacement $u_1 = 0$ is applied to prevent warping.
In order to avoid rigid body motions, the displacement is fully clamped in the plane $\min_{\ve X \in \mathcal B_0} X_1$, i.e., $\ve u = \ve 0$ holds.
Both, the basic conditions model \eqref{eq:psi_NN} and the PANN \eqref{eq:psi_corr}, have been implemented within the~FE~toolbox~\emph{FEniCS}~\cite{Alnaes2015,Logg2012}. 
Therein, the stress relations given by Eqs.~\eqref{eq:StressNN} and \eqref{eq:stress_corr} and the material tangents
\begin{align}
	\tte C^{\text{NN,}\circledcirc} := 4 \frac{\partial^2 \psi^{\text{NN,}\circledcirc}}{\partial \te C \partial \te C} \in \Lspace{4}
	\quad\text{and}\quad
	\tte C^{\text{PANN,}\circledcirc} := 4 \frac{\partial^2 \psi^{\text{PANN,}\circledcirc}}{\partial \te C \partial \te C} \in \Lspace{4} \; ,
\end{align}
which are required within the solution via a standard Newton-Raphson scheme, are calculated by means of \emph{automatic differentiation}. 

To evaluate the NNs' prediction quality, a comparison to the local stress field $P_{31}^\text{nh}$ is considered in the following, cf. Fig.~\ref{fig:results_torsion_ellipse_training}(b). We start with the models trained by complex multiaxial stress-strain states stored in the dataset $\mathcal D^{\text{red},\circledcirc}$, where the covered domain in the invariant space is shown in Fig.~\ref{fig:results_FEM_invariants}.
For both NNs, relative errors below $\SI{0.004}{\percent}$ occur, see the surface plots given in Fig.~\ref{fig:results_torsion_ellipse_training}(c), (d), i.e., the predictions are almost perfect w.r.t. the reference stresses $P_{31}^\text{nh}$. This is not surprising, since a mapping of the deformations of the torsional sample into the invariant space shows that it is almost completely covered by $\mathcal D^{\text{red},\circledcirc}$. Therefore, nearly no extrapolation is necessary here, see also the discussion in \cite{kalina2022a}.
In the next step, we consider the NN-based models only trained by uniaxial stress states stored in the dataset $\mathcal D^{\text{uniaxial},\circledcirc}$, where the covered curve in the invariant space is again shown in Fig.~\ref{fig:results_FEM_invariants}. Looking now at the results in Figs.~\ref{fig:results_torsion_uniaxial_training}(c) and (d), we see a significant error of $\SI{3.7}{\percent}$ for the basic conditions model, which is due to the need for extrapolation.
However, despite this, very low errors below $\SI{0.14}{\percent}$ are achieved with the PANN model.

Consequently, in this example, both the basic conditions model and the PANN approach are very well able to describe the learned nonlinear constitutive behavior within the FE simulation of a comparatively complex load case, whereas the PANN is better, especially if extrapolation is required.
Thereby, NNs with $N^\text{NN}=8$ and $N^\text{NN}=4$ neurons in only one hidden layer are used, respectively, which is very small compared to typical NNs applied to problems originating from computational mechanics. Moreover, the implemented basic conditions model and the PANN provide the typical quadratic convergence of Newton iteration and are therefore computationally very efficient.

\section{Conclusion}\label{sec:conc}

In the present work, an NN-based constitutive model for compressible finite strain hyperelasticity is proposed. This approach denoted as PANN fulfills all common constitutive conditions belonging to the class of hyperelasticity, i.e.,  \emph{thermodynamic consistency, symmetry of the stress tensor, objectivity, material symmetry, polyconvexity, growth condition}, as well as \emph{normalization of energy and stress}, in an exact way. 
Furthermore, the non-negativity of the neural network potentials is numerically examined by sampling the space of physically admissible deformation states. For the isotropic model, the sampling space required for that is reduced based on analytical considerations.
The proposed model allows the description of highly nonlinear hyperelastic relationships while taking into account the underlying physics and is trainable by using standard machine learning libraries such as Tensorflow.
In addition, a proof for the non-negativity of the compressible Neo-Hooke potential is presented, which, to the best of the authors' knowledge, has not been done yet.

Starting with a short literature review on NN-based elastic models, an introduction on finite strain hyperelasticity including an overview on general requirements as well as two specific anisotropy classes and models is given. Based on this, the PANN approach is built up step by step: 
using different sets of invariants as inputs for a convex neural network, the model fulfills the balance of angular momentum, objectivity and material symmetry conditions, as well as thermodynamic consistency and polyconvexity. Then, the volumetric growth condition is fulfilled by using an analytical growth term. Finally, energy and stress normalization are fulfilled by polyconvex normalization terms. The stress normalization terms depend on the material symmetry group and are exemplarily derived for isotropic and transversely isotropic material behavior. 
However, the procedure for fulfilling physical conditions, e.g., using normalization terms, can also be applied to analytical or other machine learning approaches.
Afterwards, the applicability of the PANN models is demonstrated, where a calibration to isotropic and transversely isotropic data generated with analytical potentials is performed. For all cases, even for highly multiaxial deformation states and noisy stress-strain data, a highly accurate and robust prediction quality has been shown. In addition, it has been shown that the PANN is characterized by an extremely good extrapolation capability. Finally, the straightforward application into FE simulations is demonstrated.

Summarizing, the introduced PANN model for compressible finite strain hyperelasticity has shown to be an efficient tool, which can be used in numerous applications stemming from solid mechanics. Thereby, including physics into the NN-based model is the crucial step for several reasons: first of all it leads to reliable, i.e., physically sensible, model predictions. But more than that, it is also essential in order to improve the generalization properties of NNs \cite{kumar2022,karniadakis2021} and allows for extrapolation \cite{kalina2022b,klein2022a}. 
Furthermore, only with the pronounced mathematical structure that the inclusion of constitutive conditions provides, it is possible to calibrate the models with small amounts of data which are usually available in engineering applications. Finally, this also enables the use of comparatively small network architectures, cf. \cite{kalina2022b,klein2022a}. 
Thus, when constructing NN-based constitutive models, as many constitutive conditions should be fulfilled in an exact way as possible, as this is the only way to ensure their fulfillment with absolute certainty. However, for some applications this might not be possible, e.g., when for the symmetry group under consideration no complete functional basis in invariants is available and using invariants would restrict the model flexibility too much, cf. \cite{klein2021}. Only then the structure of the model should be weakened by fulfilling some constitutive conditions in an approximate fashion, in order to gain more model flexibility. Besides that, as already mentioned, the structure and reliability that the exact fulfillment of constitutive conditions provides, should always be prioritized.

In order to generalize our proposed NN-based framework for elastic materials, several extensions are planned in the future. For instance, polyconvex normalization terms for further material symmetry groups~\cite{Ebbing2010} have to be derived. Furthermore, in order to allow an automated discovery of type and orientation of the underlying anisotropy, the usage of tensor-basis NNs would be a valuable addition~\cite{Fuhg2022b}. The application of the PANN model in the identification of material models from experimental data is also promising~\cite{flaschel2021,thakolkaran2022}. Finally, an extension to multiphysics problems~\cite{Kalina2020,klein2022a,zlatic2023} is feasible to expand the possible field of application.

\section*{Acknowledgements}
\label{sec:acknowledgements}

Dominik K. Klein and Oliver Weeger acknowledge funding from the Deutsche Forschungsgemeinschaft (DFG – German Research Foundation) – Grant No. 492770117 and support by the Graduate School of Computational Engineering within the Centre of Computational Engineering at the TU Darmstadt.

All presented computations were performed on a PC-Cluster at the Center for Information Services and High Performance Computing (ZIH) at TU Dresden.
The authors thus thank the ZIH for generous allocations of computer time.
Finally, the authors would like to thank Franz Hirsch and Philipp Metsch for providing the serverjob scripts to communicate with the HPC-Cluster.
\section*{CRediT authorship contribution statement}
\label{sec:CrediT}

\textbf{Lennart Linden:} Conceptualization, Formal analysis, Investigation, Methodology, Visualization, Software, Validation, Visualization, Writing - original draft, Writing - review and editing. 
\textbf{Dominik Klein:} Conceptualization, Formal analysis, Methodology, Visualization, Software, Validation, Writing - original draft, Writing - review and editing. 
\textbf{Karl A. Kalina:} Conceptualization, Formal analysis, Methodology, Visualization, Software, Writing - original draft, Writing - review and editing. 
\textbf{J\"org Brummund:} Formal analysis, Methodology, Writing - review and editing. 
\textbf{Oliver Weeger:} Conceptualization, Funding acquisition, Resources, Writing - review and editing.
\textbf{Markus K\"astner:} Funding acquisition, Resources, Writing - review and editing.
\section*{Declarations}
\label{sec:declarations}

\textbf{Conflict of interest:} The authors declare that they have no conflict of interest.

\appendix
\section{Multilayered neural networks}
\label{sec:multilayer}

\begin{figure}[t!]
	    \centering
	    \includegraphics[width=\textwidth]{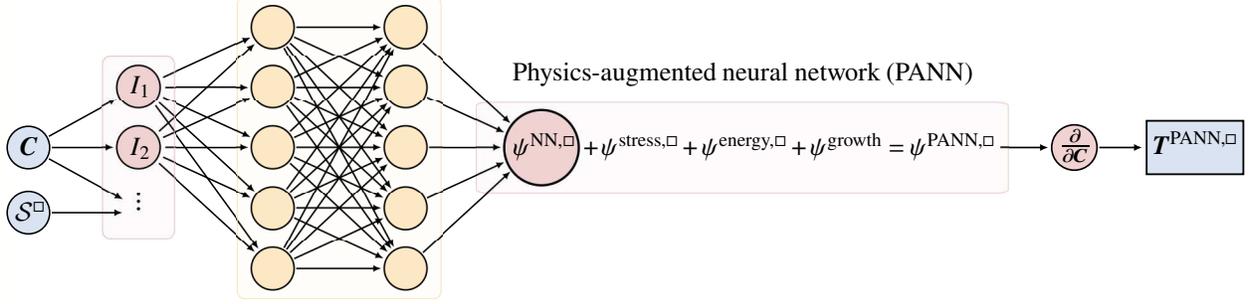}
	    \caption{Illustration of the multilayered PANN based constitutive model for the material symmetry group $\square$ under consideration.
	    }
	    \label{fig:PANN_scheme_multiple_hidden_layer}
	\end{figure}

In this work, sets of invariants are used as inputs for FFNNs \cite{aggarwal2018,kollmannsberger2021} with scalar-valued output, where the output is used to model a hyperelastic potential. In Sec.~\ref{sec:model}, a FFNN architecture with only one hidden layer is introduced, which prooves to be flexible enough for a lot of practical applications \cite{klein2022a,kalina2022b}. 
Nevertheless, the methods introduced in Sec.~\ref{sec:model} are not restricted to network architectures containing only one hidden layer, and for the sake of completeness, multilayered network architectures, cf. Fig.~\ref{fig:PANN_scheme_multiple_hidden_layer}, are now introduced.

In a nutshell, FFNNs can be seen as a composition of multiple vector-valued functions, where the components are referred to as nodes or neurons and the functions acting in each node are referred to as activation functions. FFNNs can gain flexibility in two ways: either the number of nodes in each hidden layer is increased, as it is done for one hidden layer in Sec.~\ref{sec:model}, or the amount of hidden layers is increased.
Generalizing the single-layered architecture of Eq.~\eqref{eq:psi_NN} for a network architecture with $H$ hidden layer and $N^{\text{NN},H}$ nodes in each hidden layer yields
	\begin{align}
	\label{eq:psi_NN_ML}
    \te A^{[1]} _\alpha &=  
   \mathscr{F}\Big(\sum_{\beta=1}^{m} w_{\alpha\beta}^{[1]} I_\beta + \sum_{\gamma=1}^{A} w_{\alpha\gamma}^{*[1]} I_\gamma^* + b_\alpha^{[1]}\Big)\in\R^{N^{\text{NN},1}}\; ,
    \\
    \te A_\alpha ^{[h]} &=\mathscr{F}\Big(\sum_{\beta=1}^{N^{\text{NN},h-1}}w_{\alpha\beta}^{[h]} \te A_\beta^{[h-1]}+b_\alpha^{[h]} \Big)\in\R^{N^{\text{NN},h}}\; \text{with } h=2,\dotsc,H\;,
    \\
    \psi^{\text{NN,}\square}(\ve{\mathcal I}^*)&= \sum_{\alpha=1}^{N^{\text{NN},H}} W_{\alpha}\,\te A^{[H]} _\alpha\in\R\; ,
	\end{align}
	with the polyconvex, irreducible and independent invariants $I_\beta(\te C, \mathcal S^\square)$ as well as the additional invariants $I_\gamma^*(\ve{\mathcal I})$ as defined in Sec.~\ref{sec:model}. By the special choice of activation function $\mathscr{F}$ as convex and non-decreasing in every hidden layer and $W_\alpha\geq 0$, the polyconvexity of the invariants is preserved. Again, the \emph{Softplus} activation function $\Softplus(x):=\log(1+\exp(x))$ is applied, which is convex and non-decreasing for non-negative weights and arbitrary bias values, and overall the conditions
\begin{align}
	\label{eq:weights_poly_multilayer}
    W_\alpha, w_{\alpha\beta}^{[h]}, w_{\alpha\gamma}^{*[1]} \in \R_{\geq 0}, b_\alpha \in\R \quad \forall h\in\N_{\leq H}, \alpha\in\N_{\leq N^{\text{NN},h-1}}, \beta \in \N_{\leq m}, \gamma \in \N_{\leq A}\, 
	\end{align}
result in a polyconvex neural network, cf. Remark~\ref{rem:ICNN}, see also \cite{klein2021} for a more extensive discussion and explicit proofs.

\section{Derivatives of invariants}
\label{sec:derivatives_inv}

\renewcommand{\arraystretch}{2}
\begin{table}[t!]
\centering
\caption{Derivatives of isotropic and transversely isotropic invariants.}
\begin{small}
\begin{tabular}{| l | l |c|c| } 
\hline
 \cellcolor[gray]{.7} & \cellcolor[gray]{.9} $I_{\alpha}$ & \cellcolor[gray]{.9} $\dfrac{\partial I_{\alpha}}{\partial \te C}$
 \\[1ex]
\hline
\multirow{3}{5em}{Isotropic} 

 \cellcolor[gray]{.9} & $I_1 := \tr \te C$ & $\te 1$ \\ 
 \cellcolor[gray]{.9} & $I_2 := \tr(\cof \te C)$ & $I_1 \te 1 - \te C$ \\ 
 \multirow{-2.8}{6em}{\cellcolor[gray]{.9}Isotropy} & $I_3 := \det \te C$ & $\cof \te C$  
\\[1ex]
\hline
\cellcolor[gray]{.9}
 & $I_4 := \tr(\te C\cdot\te G)$ & $\te G$ \\ 
\multirow{-2}{6em}{\cellcolor[gray]{.9}Transverse isotropy} & $I_5 := \tr(\cof(\te C) \cdot \te G)$ & $I_5\te C^{-1}- \cof(\te C) \cdot\te G \cdot \te C^{-1}$ 
\\[1ex]
\hline
\end{tabular}
\end{small}
\label{tab:invs}
\end{table}
For the convenience of the reader, in Tab.~\ref{tab:invs}, both isotropic and transversely isotropic invariants, as well as their derivatives w.r.t. $\te C$, are provided.
From this, the derivative of the determinant $J=\sqrt{I_3}$ follows as
\begin{align}
    \frac{\partial J}{\partial \te C}=\frac{J}{2}\te C^{-1}\;,
\end{align}
which implies the derivative of the adapted invariant $I_1^* := -2 J$  as
\begin{align}
     \frac{\partial I_1^*}{\partial \te C}&=-J\te C^{-1}\,.
\end{align}

\section{Non-negativity of the strain energy density}\label{app:non_neg}
In this appended section, the non-negativity of isotropic hyperelastic energy functions is discussed by analytical consideration. In many cases strain energy density functions are formulated in terms of invariants, but it should be noted that not all arbitrary states in the space of invariants are also physically admissible, since these can lead to real negative or even complex principal stretches. 
Consequently, it is necessary to determine restrictions for the invariants in the following, which ensure positive definiteness of the right Cauchy-Green deformation tensor.
Then, for certain strain energy density functions, it can be proven that they only lead to non-negative values for physically admissible deformation states. This procedure will be shown in the following.

Let $I_1 := \tr{\te C}, \; I_2 := \tr{(\cof \te C)}$ and $I_3 := \det \te C$ be the principal invariants of the right Cauchy-Green deformation tensor $\te C:=\te F^T \cdot \te F \in \Sym$.
We denote the eigenvalues of the right Cauchy-Green deformation tensor as $\kappa_1, \kappa_2$ and $\kappa_3$.
These eigenvalues can be represented by the principal stretches $\lambda_1, \lambda_2$ and $\lambda_3$, which are the eigenvalues of the deformation gradient $\te F \in \GL^+(3)$, i.e., $\kappa_\alpha = \lambda_\alpha^2$ holds for all $\alpha \in \N_{\leq 3}$.
\newtheorem{Definition}{Definition}[section]
\begin{Definition}
A \textit{deformation state} is called \textit{physically admissible} if the deformation tensor $\te C \in \Sym$ is positive definite, i.e., if all associated eigenvalues are positive $(\kappa_1, \kappa_2, \kappa_3 > 0)$.
\end{Definition}
\newtheorem{Lemma}[Definition]{Lemma}
\newtheorem{Theorem}[Definition]{Theorem}
\newtheorem{Corollary}[Definition]{Corollary}
\begin{Lemma}
\label{lemma:characteristic_equation}
The characteristic equation 
\begin{align}
	\label{eq:eigenvalues}
    \kappa^3 - I_1\kappa^2 + I_2 \kappa - I_3 = 0\; ,
\end{align}
of the right Cauchy-Green deformation tensor $\te C \in \Sym$ has real eigenvalues, if and only if the restriction
\begin{align}
	\label{eq:restrictions_eigenvalues}
    \Gamma := \Big(\frac{q}{2}\Big)^2 + \Big(\frac{p}{3}\Big)^3 \leq 0\; , \quad \text{ with }\; p := I_2 - \frac{I_1^2}{3}\; , \; q := -\frac{2 I_1^3}{27} + \frac{I_1 I_2}{3} - I_3
\end{align}
is fulfilled.
\begin{proof}
According Cardano's formula, one obtains real solutions of the cubic Eq.~\eqref{eq:eigenvalues}, if and only if the restriction~\eqref{eq:restrictions_eigenvalues} is fulfilled~\cite{mathhandbook2007}.
\end{proof}
\end{Lemma}
\begin{Theorem}
\label{theorem:admissible_states}
A \textit{deformation state} is \textit{physically admissible}, if and only if the following restrictions 
\begin{align}
	\label{eq:boundary_condition}
    \Gamma(I_1, I_2, I_3) = \frac{1}{108} \Big( 4I_1^3 I_3 - I_1^2 I_2^2 + 4 I_2^3 + 27 I_3^2 - 18 I_1 I_2 I_3\Big) \leq 0\quad \text{and} \quad I_1, I_2, I_3 > 0
\end{align}
are fulfilled.
\begin{proof}
$"\Leftarrow":$ Given that restrictions~\eqref{eq:boundary_condition} are satisfied and since the invariants $I_1, I_2, I_3 > 0$ are positive, it can be shown by a proof of contradiction that the eigenvalues $\kappa_\alpha$ are positive real solutions of Eq.~\eqref{eq:eigenvalues}.
The invariants are given by $I_1 = \kappa_1 + \kappa_2 + \kappa_3$, $I_2 = \kappa_1 \kappa_2 + \kappa_1 \kappa_3 + \kappa_2 \kappa_3$ and $I_3 = \kappa_1 \kappa_2 \kappa_3$ and it is assumed that Eq.~\eqref{eq:eigenvalues} is satisfied. 
The invariant $I_3$ is non-negative exactly if none or exactly two eigenvalues $\kappa_\alpha$ are negative, since the eigenvalues are real numbers according Lemma~\ref{lemma:characteristic_equation}.
Furthermore, the condition $I_3 \neq 0$ is equivalent to $\kappa_\alpha \neq 0$ for all $\alpha \in \N_{\leq 3}$.
Hence, only the case that two eigenvalues are negative and one is positive  must be investigated.  
Without loss of generality let $\kappa_2$ and $\kappa_3$ be negative and $\kappa_1$ be positive $(\kappa_1 > 0, \kappa_2 < 0, \kappa_3 < 0)$.
Then, we obtain
\begin{align}
   I_2 = \kappa_1 (\kappa_2 + \kappa_3) + \kappa_2 \kappa_3 &> 0 \; ,\\
   \Leftrightarrow \frac{\kappa_2 \kappa_3}{-(\kappa_2 + \kappa_3)} &> \kappa_1 > -(\kappa_2 + \kappa_3)\; ,
\end{align}
where in the last inequality the relation $I_1 = \kappa_1 + \kappa_2 + \kappa_3 > 0$ was used.
After multiplying with the term $-(\kappa_2 + \kappa_3) > 0$, this leads to the contradiction
\begin{align}
0 > \kappa_2^2 + \kappa_3^2 + \kappa_2 \kappa_3\; ,
\end{align}
because the right hand side only consists of positive terms.
\\
$"\Rightarrow":$ This follows directly from the definitions $I_1 = \kappa_1 + \kappa_2 + \kappa_3$, $I_2 = \kappa_1 \kappa_2 + \kappa_1 \kappa_3 + \kappa_2 \kappa_3$ and $I_3 = \kappa_1 \kappa_2 \kappa_3$ as well as from Cardano's formula~\cite{mathhandbook2007}.
\end{proof}
\end{Theorem}
\begin{Lemma}
\label{lemma:equal_eigenvalues}
Let the restrictions~\eqref{eq:boundary_condition} be fulfilled. If and only if the case $\Gamma = 0$ holds, then at least \textit{two eigenvalues} $\kappa_\alpha$ are \textit{equal}.
The single eigenvalue will be denoted as $x > 0$ and the double eigenvalue as $y > 0$.
As a consequence, the invariants result in
\begin{align}
	\label{eq:invariants_eigenvalues}
    I_1 = x +2y > 0\ , \quad I_2 = 2xy + y^2 > 0\ , \quad I_3 = xy^2 > 0\; .
\end{align}
Then, the partial derivatives are given by
\begin{align}
	\label{eq:derivative_invariants_eigenvalues_1}
    \frac{\partial I_1}{\partial x} &= 1 > 0\; , & \frac{\partial I_1}{\partial y} &= 2 > 0 \; , \\
    \label{eq:derivative_invariants_eigenvalues_2}
    \frac{\partial I_2}{\partial x} &= 2y > 0\; , & \frac{\partial I_2}{\partial y} &= 2x + 2y > 0\; ,\\
    \label{eq:derivative_invariants_eigenvalues_3}
    \frac{\partial I_3}{\partial x} &= y^2 > 0\; , & \frac{\partial I_3}{\partial y} &= 2xy > 0\; .
\end{align}
\begin{proof}
The first result follows directly from Cardano's formula~\cite{mathhandbook2007}. Then, the partial derivatives are basic calculations.
\end{proof}
\end{Lemma}

\subsection{Neo-Hooke model}

\begin{Theorem}
\label{theorem:neo_hooke}
The strain energy density function
\begin{align}
\label{eq:appendix_energy_NH}
    \psi^{\text{nh}}(I_1,I_3)=\frac{1}{2}\left(\mu\left(I_1-\ln I_3 - 3\right)+\frac{\lambda}{2}\left(I_3-\ln I_3 -1\right)\right)\,, \quad \mu, \lambda > 0
\end{align}
of the isotropic \textit{Neo-Hooke model} is \textit{non-negative} for all physically admissible deformation states.
\begin{proof}
The continuously differentiable function $\psi^\text{nh}: \R^3 \to \R, (I_1, I_2, I_3) \mapsto \psi^\text{nh}(I_1, I_3)$ is convex w.r.t. the principal invariants $I_1, I_2, I_3$ and has no local extremum, since the gradient w.r.t. the principal invariants does not vanish.
Note that this Neo-Hooke model does not depend on the principal invariant $I_2$ explicitly.
Likewise, the gradient does not vanish if the domain of the function $\psi^\text{nh}(I_1, I_3)$ is restricted to the subset $\Omega := \{(I_1, I_2, I_3) \in \R^3 \mid \Gamma(I_1, I_2, I_3) \leq 0, \ I_1, I_2, I_3 > 0\}$ according Theorem~\ref{theorem:admissible_states}.
Accordingly, the minimum value is assumed to be on the boundaries of the admissible domain $\Omega$ given by
\begin{align}
	\label{eq:bound_gamma}
    \partial\Omega_\Gamma &:= \{(I_1, I_2, I_3) \in \R^3 \mid \Gamma(I_1, I_2, I_3) = 0, \ I_1 > 0, I_2 > 0, I_3 > 0\}\;,\\
    \label{eq:bound_1}
    \partial\Omega_1 &:= \{(I_1, I_2, I_3) \in \R^3 \mid \Gamma(I_1, I_2, I_3) \leq 0, \ I_1 = 0, I_2 > 0, I_3 > 0\}\;,\\
    \label{eq:bound_2}
    \partial\Omega_2 &:= \{(I_1, I_2, I_3) \in \R^3 \mid \Gamma(I_1, I_2, I_3) \leq 0, \ I_1 > 0, I_2 = 0, I_3 > 0\}\;,\\
    \label{eq:bound_3}
    \partial\Omega_3 &:= \{(I_1, I_2, I_3) \in \R^3 \mid \Gamma(I_1, I_2, I_3) \leq 0, \ I_1 > 0, I_2 > 0, I_3 = 0\}\;.
\end{align}
First, we note that for $I_{1}\rightarrow 0^+$ or $I_{2}\rightarrow 0^+$, also $I_3\rightarrow 0^+$ holds, due to the fact that the principal invariants are real positive values $(I_1, I_2, I_3 > 0)$. Since the term $-\ln I_3$ is dominating according Eq.~\eqref{eq:appendix_energy_NH}, $\partial\Omega_{1,2,3}$ leads to $\lim_{I_3\to 0^+} \psi^\text{nh}(I_1,I_3)=\infty$, which is not a local minimum. 

Second, we consider $\partial\Omega_\Gamma$. 
Consequently, we now consider only the independent variables $x, y \in \R_{> 0}$ according to Lemma~\ref{lemma:equal_eigenvalues}.
To find an extremum of $\psi^{\text{nh}}$ on $\partial\Omega_\Gamma$, the partial derivatives of the potential w.r.t. the independent variables are computed and set to zero: 
\begin{align}
	\label{eq:derivative_neo_hooke_1}
    \frac{\partial \psi^{\text{nh}}}{\partial x} = \frac{\partial \psi^{\text{nh}}}{\partial I_1}\frac{\partial I_1}{\partial x} + \frac{\partial \psi^{\text{nh}}}{\partial I_3}\frac{\partial I_3}{\partial x} 
    = \frac{\mu}{2} \left(1 - \frac{1}{x}\right) + \frac{\lambda}{4}\left(1 - \frac{1}{xy^2}\right)y^2 = 0\; ,\\
    \label{eq:derivative_neo_hooke_2}
    \frac{\partial \psi^{\text{nh}}}{\partial y} = \frac{\partial \psi^{\text{nh}}}{\partial I_1}\frac{\partial I_1}{\partial y} + \frac{\partial \psi^{\text{nh}}}{\partial I_3}\frac{\partial I_3}{\partial y} 
    = \mu \left(1 - \frac{1}{y}\right) + \frac{\lambda}{2}\left(1 - \frac{1}{xy^2}\right)xy = 0\; ,
\end{align}
where Eqs.~\eqref{eq:derivative_invariants_eigenvalues_1}--\eqref{eq:derivative_invariants_eigenvalues_3} with $x, y\neq 0$ are used.
By subtracting the half of Eq.~\eqref{eq:derivative_neo_hooke_2} from Eq.~\eqref{eq:derivative_neo_hooke_1} we obtain the equation
\begin{align}
	\label{eq:derivative_neo_hooke_mod}
   \frac{1}{2}(y-x) \left(-\frac{\mu}{xy} + \frac{\lambda}{2}\left(y - \frac{1}{xy}\right)\right) = 0\; ,
\end{align}
which obviously holds for $x = y$.
The other solution can be obtained from
\begin{align}
-\mu + \frac{\lambda}{2}\left(xy^2 - 1\right) = 0 \quad \Leftrightarrow \quad y^2 = \frac{1}{x}\left(\frac{2\mu}{\lambda} + 1\right)\; ,
\end{align}
but when substituted into Eq.~\eqref{eq:derivative_neo_hooke_1}, this implies the contradiction $\mu = 0$.

Finally, we substitute the solution $x=y$ into Eq.~\eqref{eq:derivative_neo_hooke_1}, which leads to the cubic equation
\begin{align}
x^3 + \frac{2\mu}{\lambda} x - \left(\frac{2\mu}{\lambda} + 1\right) = 0\; ,
\end{align}
which has obviously the solution $x = 1$.
Using polynomial division, we can obtain the quadratic equation
\begin{align}
x^2 + x \left(\frac{2\mu}{\lambda} + 1\right) = 0\; ,
\end{align}
which does not lead to real solutions, since $\mu, \lambda > 0$ hold.
In fact, the function $\psi^\text{nh}(x)$ is convex w.r.t. the triple eigenvalue $x\in(0,\infty)$ of the volumetric deformation states $\te C=x \te 1$, since $\frac{\partial^2 \psi^\text{nh}}{\partial x^2}>0$ holds for all positive triple eigenvalues of the spherical tensor.

Consequently, $x = y = 1$ is a global minimum point of $\psi^\text{nh}(x,y)$ and thus the unique real and non-negative solution of Eqs.~\eqref{eq:derivative_neo_hooke_1} and \eqref{eq:derivative_neo_hooke_2} on the admissible set $\Omega$. Finally, $\psi^{\text{nh}}(x=1,y=1) = 0$ can be determined, i.e., the undeformed state has zero energy.
Hence, $\psi^{\text{nh}}(I_1,I_3) \geq 0$ holds for all physically admissible deformation states.
\end{proof}
\end{Theorem}
\subsection{Isotropic PANN}
\begin{Theorem}
\label{theorem:isotropic_PANN}
Consider the \textit{isotropic PANN model}
\begin{align}
	\label{eq:apendix_isotropic_PANN}
    \psi^{\text{PANN,}\circledcirc}(\ve{\mathcal I}^*) = \sum_{\alpha=1}^{N^\text{NN}} W_{\alpha}\,\Softplus\left(\sum_{\beta=1}^{3} w_{\alpha\beta} I_\beta + w_{\alpha 1}^* I_1^* + b_\alpha\right) + \Big(J + \frac{1}{J} - 2\Big)^2 - \mathfrak{n}(J-1) + \psi^{\text{energy,}\circledcirc}\; ,
\end{align}
including the normalization constant
\begin{align}
\label{eq:appendix_normalization_A_iso}
\mathfrak{n} := 2 \,\bigg(\,\frac{\partial \psi^{\text{NN},\circledcirc}}{\partial I_1}
 +
 2\frac{\partial \psi^{\text{NN},\circledcirc}}{\partial I_2}
+
 \frac{\partial \psi^{\text{NN},\circledcirc}}{\partial I_3}
+\frac{\partial \psi^{\text{NN},\circledcirc}}{\partial I_1^*} \frac{\partial I_1^*}{\partial I_3}
 \bigg)\Bigg\rvert_{\te C = \te 1} \in \R\; ,
\end{align}
with $\ve{\mathcal I}^* := (I_1, I_2, I_3, I_1^*)$ and the additional invariant $I_1^* := -2\sqrt{I_3} = -2J$. Given that
\begin{align}
\label{eq:condition_isotropic_PANN}
\frac{\partial \psi^{\text{PANN,}\circledcirc}}{\partial I_1} > 0 \quad \text{and} \quad \frac{\partial \psi^{\text{PANN,}\circledcirc}}{\partial I_2} > 0
\end{align}
holds, the potential $\psi^{\text{PANN,}\circledcirc}(\ve{\mathcal I}^*)$ can possess \textit{local minima} only for \textit{volumetric deformation states} $\te C = \lambda^2 \te 1$ with $\lambda \in (0, \infty)$.
\begin{proof}
The continuously differentiable function $\psi^{\text{PANN,}\circledcirc}: \R^3 \to \R, (I_1, I_2, I_3) \mapsto \psi^{\text{PANN,}\circledcirc}(I_1, I_2, I_3, -2\sqrt{I_3})$ is convex w.r.t. the principal invariants $I_1, I_2, I_3$ and has no local extremum, since the gradient w.r.t. the invariants does not vanish due to condition \eqref{eq:condition_isotropic_PANN}.
It should be noted that the assumptions~\eqref{eq:condition_isotropic_PANN} are not too restrictive for the approximation quality of the neural network, since they are already satisfied by small positive weights $w_{\alpha 1}, W_{\alpha}$ and $w_{\beta 2}, W_{\beta}$ for at least one $\alpha, \beta \in \N_{\leq N^\text{NN}}$.
Thus, the gradient does not vanish when the domain of the function $\psi^{\text{PANN,}\circledcirc}(\ve{\mathcal I}^*)$ is restricted to the  subset of admissible invariants $\Omega := \{(I_1, I_2, I_3) \in \R^3 \mid \Gamma(I_1, I_2, I_3) \leq 0, \ I_1, I_2, I_3 > 0\}$, as given in according Theorem~\ref{theorem:admissible_states}.

Since the growth term~\eqref{eq:GrowthNNmod} is included in the potential $\psi^{\text{PANN,}\circledcirc}(\ve{\mathcal I}^*)$ and all other terms are finite, the boundaries $\partial\Omega_1, \partial\Omega_2, \partial\Omega_3$ from Eqs.~\eqref{eq:bound_1}--\eqref{eq:bound_3} do not need to be considered. The argumentation is analogous to that in the proof of Theorem~\ref{theorem:neo_hooke}.
Consequently, only the boundary $\partial\Omega_\Gamma$ needs to be considered, given in Eq.~\eqref{eq:bound_gamma}.
The partial derivatives of the polyconvex NN-based model w.r.t.~the invariants are denoted by
\begin{align}
    \psi_{\gamma} := \frac{\partial \psi^{\text{PANN,}\circledcirc}}{\partial I_\gamma}\; .
\end{align}
In order to compute local extrema, the partial derivatives w.r.t.~the eigenvalues, which are expressed with the help of Lemma \ref{lemma:equal_eigenvalues}, are set to zero:
\begin{align}
\label{eq:derivative_pann_1}
    \frac{\partial \psi^{\text{PANN,}\circledcirc}}{\partial x} &= \psi_1 + 2y \;\psi_2 + y^2 \;\psi_3 = 0\; ,\\
    \label{eq:derivative_pann_2}
    \frac{\partial \psi^{\text{PANN,}\circledcirc}}{\partial y} &= 2 \;\psi_1 + 2(x + y) \;\psi_2 + 2xy \;\psi_3 = 0\ .
\end{align}
Subtracting the half of Eq.~\eqref{eq:derivative_pann_2} from Eq.~\eqref{eq:derivative_pann_1} yields 
\begin{align}
	\label{eq:derivative_pann_mod}
   (y-x) \left(\psi_2 + y \;\psi_3\right) = 0\; .
\end{align}
Substituting the first solution $\psi_2 = -y \;\psi_3$ into Eq.~\eqref{eq:derivative_pann_1} leads to the contradiction $\psi_1 = y^2 \;\psi_3$ and $\psi_2 =- y \;\psi_3$, which can not be fulfilled by $y > 0$ as well as the assumptions $\psi_1 > 0$ and $\psi_2 > 0$.
Consequently, the other solution $x = y$ is the only solution of the Eqs.~\eqref{eq:derivative_pann_1} and~\eqref{eq:derivative_pann_2}.

Same as for the Neo-Hooke model above, the search for a global minimum of $\psi^{\text{PANN,}\circledcirc}(\ve{\mathcal I}^*)$ thus can be narrowed down to the (convex) set of real positive triple eigenvalues $x\in (0, \infty)$ of the spherical tensor $\te C=x \te 1$.
In particular, $x= 1$ is a local minimum point by construction with $\psi^{\text{PANN,}\circledcirc}(x=1)=0$, since the normalization constant $\mathfrak{n}$ of the normalization term, given in Eq.~\eqref{eq:normalization_isotropic}, was chosen accordingly.
However, since $J=x^{3/2}$ is strictly concave in $x\in (0, \infty)$ and may appear with a positive or negative factor due the normalization, it can generally not be shown that $\psi^{\text{PANN,}\circledcirc}(x)$ is convex w.r.t. the triple eigenvalue. 
Thus, the minimum point $x=1$ may not be unique and further triple eigenvalues $x \neq 1$ of the spherical tensor $\te C=x \te 1$ might exist, for which the derivative $\frac{\partial \psi^{\text{PANN,}\circledcirc}}{\partial x}$ vanishes and $\psi^{\text{PANN,}\circledcirc}(x)< 0$ holds.
\end{proof}
\end{Theorem}
\section{Stochastic}
\label{sec:stochastic}

\begin{figure}[t!]
	    \centering
	    \includegraphics[width=\textwidth]{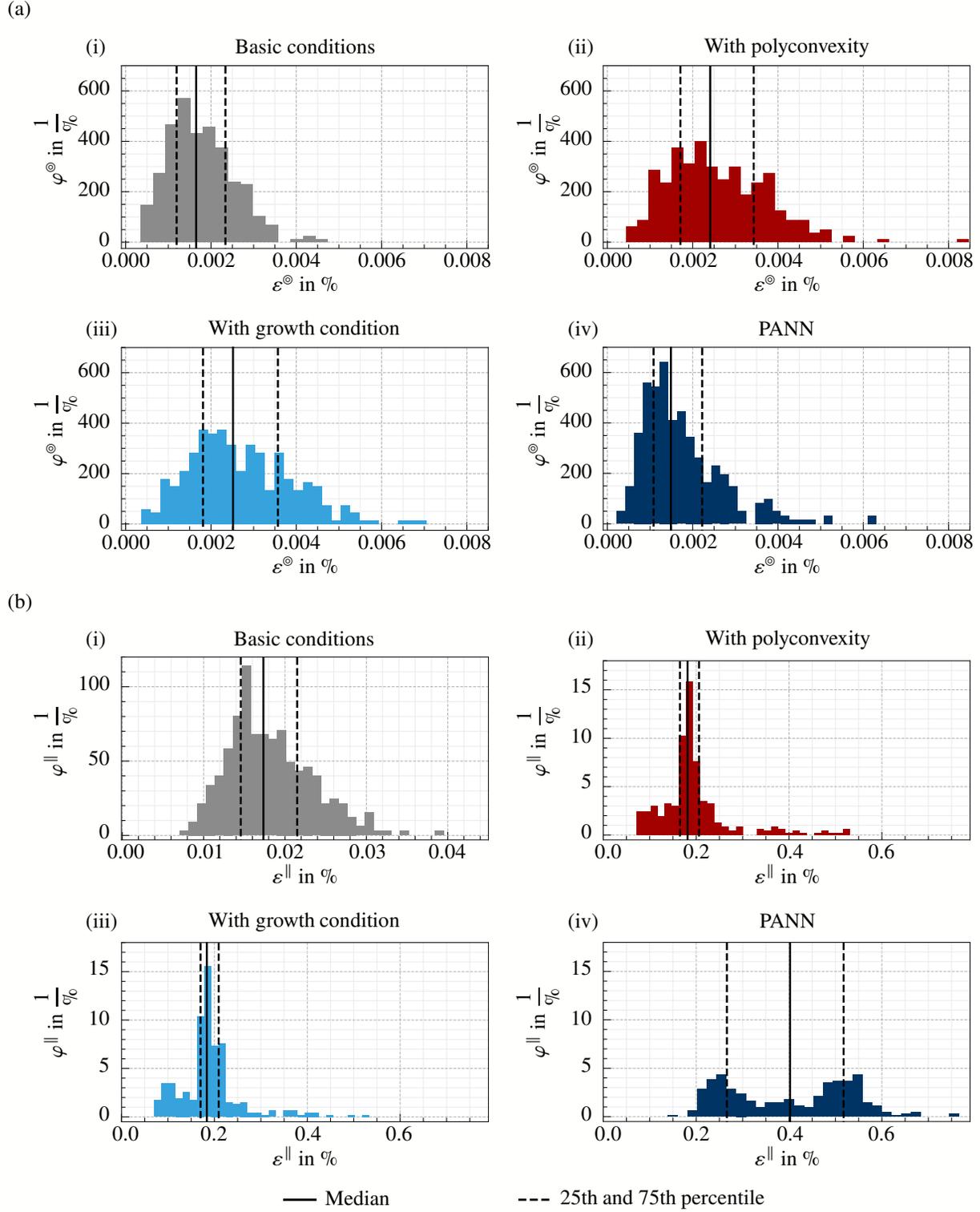}
	    \caption{Histograms with $N^\text{bins} := 30$ of the error measure $\varepsilon^\square$ given in Eq.~\eqref{eq:error_relative} for different anisotropies: (a) isotropic and (b) transversely isotropic invariant-based models. (i) basic conditions, (ii) polyconvexity, (iii) growth condition and polyconvexity, as well as (iv) PANN satisfying all conditions including normalization.
	    The results were generated with 300 training runs, selecting the best of 30 trains each. Calibration has been done with the reduced data set $\mathcal D^{\text{red},\square}$.
	    }
	    \label{fig:frobenius_error_both}
	    \end{figure}

In this appended section, the raw data obtained in the statistical study which is discussed in Sec.~\ref{subsubsec:prediction_quality} are presented. The histograms including median and 25th as well as 75th percentile for the isotropic ($\square = \circledcirc$) and the transversely isotropic ($\square = \;\parallel$) case are given in Fig.~\ref{fig:frobenius_error_both}(a)--(b), respectively.
For each model in this statistical investigation, a total of 300 training runs have been completed.
Based on the reduced data set $\mathcal D^{\text{red},\square}$, the respective NN is trained 30 times in one training run, and the parameters of the optimal training state with the lowest MSE~\eqref{eq:mse_calibration}, as described in Sec.~\ref{subsec:model_calibration}, are stored.

\clearpage
\newpage

\renewcommand*{\bibfont}{\footnotesize}
\printbibliography

\end{document}